%% file: WL_ABJM_arxiv.tex
\documentclass[11pt]{article}
\usepackage{fullpage}

\usepackage{amsmath,amssymb}
\usepackage{hyperref}
\usepackage{graphicx}

\usepackage[utf8]{inputenc}
\DeclareUnicodeCharacter{00A0}{~}

\numberwithin{equation}{section}

\input{defs}

\begin{document}

\input{title}

\input{sec1}

\input{sec2}

\input{sec3}

\input{sec4}

\input{sec5}

\input{sec6}

\section*{Acknowledgments}
We thank Cat Whiting for discussions and Jim Liu for various important clarifications. 
L.~PZ is grateful to the University of Naples ``Federico II'', Italy, for hospitality.  
The work of W.~M.\ was supported in part by the INFN, research initiative STEFI.

\begin{appendix}
\input{appA}

\input{appB}

\input{appC}

\end{appendix}

\clearpage

\input{WL_ABJM_arxiv.bbl}

\end{document}

%% file: defs.tex
\newcommand{\cf}{\mathcal{F}}

\def\eqa{\begin{eqnarray}}
\def\eqae{\end{eqnarray}}
\def\eq{\begin{equation}}
\def\eqe{\end{equation}}
\def\be{\begin{equation}}
\def\ee{\end{equation}}
\def\bea{\begin{eqnarray}}
\def\eea{\end{eqnarray}}
\def\ba{\begin{array}}
\def\ea{\end{array}}
\def\bd{\begin{displaymath}}
\def\ed{\end{displaymath}}

\def\tr{{\rm tr}}
\def\>{\rangle}
\def\<{\langle}

\newcommand{\adsfour}{{\mathrm{AdS_4}}}
\newcommand{\adstwo}{{\mathrm{AdS_2}}}
\newcommand{\cpspace}{\mathbb{C}P}
\newcommand{\cpthree}{{\cpspace^3}}

\newcommand{\Dsix}{{\mathrm{D6}}}
\newcommand{\Dtwo}{{\mathrm{D2}}}
\newcommand{\Toneone}{\tilde{T}^{1,1}}

\newcommand{\calE}{{\mathcal{E}}}

\newcommand{\flatind}[1]{{\underline{#1}}}

\newcommand{\ie}{i.e.,\ }

\newcommand{\e}[1]{\operatorname{e}^{#1}}

\newcommand{\tn}{\tilde{\nabla}}

\newcommand{\cll}{C_{j,l}}

\newcommand{\comm}[2]{\left[#1,#2\right]}

\newcommand{\slD}{\slash \mspace{-12.0mu} D}

%% file: title.tex
\begin{titlepage}
\hfill MCTP-16-14
\vspace{3em}
\begin{center}

{\Large \textbf{Spectra of Certain Holographic ABJM Wilson Loops}\\[1em]
\textbf{in Higher Rank Representations}}\\[2em]

\renewcommand{\thefootnote}{\fnsymbol{footnote}}%
Wolfgang M{\"u}ck${}^{a,b}$\footnote[1]{wolfgang.mueck@na.infn.it}, %
Leopoldo A.~Pando Zayas${}^{c,d}$\footnote[2]{lpandoz@umich.edu} and %
Vimal Rathee${}^{d}$\footnote[3]{vimalr@umich.edu}.\\[2em]%
\renewcommand{\thefootnote}{\arabic{footnote}}%
${}^a$\emph{Dipartimento di Fisica ``Ettore Pancini'', Universit\`a degli Studi di Napoli ``Federico II''\\ Via Cintia, 80126 Napoli, Italy}\\[1em] 
${}^b$\emph{Istituto Nazionale di Fisica Nucleare, Sezione di Napoli\\ Via Cintia, 80126 Napoli, Italy}\\[1em]
${}^c$\emph{The Abdus Salam International Centre for Theoretical Physics\\ Strada Costiera 11,  34014 Trieste, Italy}\\[1em]
${}^d$\emph{Michigan Center for Theoretical Physics,  Randall Laboratory of Physics\\ The University of
Michigan,  Ann Arbor, MI 48109, USA}\\[2em]

\abstract{The holographic configurations dual to Wilson loops in higher rank representations in the ABJM theory are described by branes with electric flux along their world volumes. In particular,  D2 and D6 branes with electric flux play a central role as potential dual to totally symmetric and totally antisymmetric representations, respectively. 
We compute the spectra of excitations of these brane configurations in both, the bosonic and fermionic sectors. We highlight a number of aspects that distinguish these configurations from their D3 and D5 cousins including new peculiar mixing terms  in the fluctuations.  We neatly organize the spectrum of fluctuations into the corresponding supermultiplets.}

\end{center}

\end{titlepage}

\tableofcontents
\clearpage

%% file: sec1.tex
\section{Introduction}
\label{intro}
The AdS/CFT correspondence conjectures a mathematical equivalence between string theories and gauge theories. Generically, when the field theory is strongly coupled, the string theory description is weakly coupled and reduces to supergravity. Naturally, most of the explorations have been centered in understanding strong coupling gauge theory phenomena using a weakly coupled gravity description enhanced with classical strings and branes in the corresponding supergravity backgrounds. Going beyond the supergravity limit, that is, solving the full string theory in curved spacetimes, such as $AdS_5\times S^5$ or $\adsfour\times\cpthree$, with Ramond-Ramond fluxes presents a formidable challenge. Given these technical difficulties, it would be particularly illuminating to use the AdS/CFT duality to understand quantum aspects of string perturbation theory in these situations. 

The field theory side of the correspondence has recently provided a plethora of exact results by means of supersymmetric localization. For example, in   $\mathcal{N}=4$ Supersymmetric Yang-Mills (SYM) and in a $\mathcal{N}=6$ Cherns-Simons theory known as ABJM \cite{Aharony:2008ug}, which are the field theory duals of string theories on $AdS_5\times S^5$ and $\adsfour\times\cpthree$, respectively, exact expressions for the vacuum expectation value of some supersymmetric Wilson loops have been obtained \cite{Pestun:2007rz,Klemm:2012ii}. These exact results provide fertile ground to explore the holographic side beyond the leading semi-classical approximation and, in particular, generate guidance as how to enhance string perturbation theory by pointing to crucial missing elements in the standard semi-classical quantization approach.

Indeed, there has been  a concerted effort toward matching the holographic one-loop corrections with subleading terms in the field theory side  \cite{Forste:1999qn,Drukker:2000ep,Sakaguchi:2007ea,Kruczenski:2008zk,Kristjansen:2012nz}. More recently, in an attempt to tame  some of the intrinsic ambiguities on the holographic side, the ratio of $1/4$ and $1/2$ BPS Wilson loops has been compared to the field theory ratio \cite{Forini:2015bgo,Faraggi:2016ekd} yielding some improvement in the comparison and pointing to interesting aspects of  string perturbation theory.  One of our main motivations is to continue to construct ever more stringgent tests that will clarify the nature and aspects of the corrections.

There is also an ongoing program of extending one-loop corrections to holographic configurations dual to Wilson loops in higher rank representations of the $SU(N)$  gauge group  in $\mathcal{N}=4$ SYM \cite{faraggi:2011bb,Faraggi:2011ge,Buchbinder:2014nia,Faraggi:2014tna}. 

In this manuscript we take a step towards the understanding, beyond the leading order, of holographic configurations that are expected to correspond to supersymmetric Wilson loops in higher rank representations \cite{Drukker:2008zx} in the ABJM theory. Namely, we construct the spectra of quantum fluctuations of a D6 and a D2 brane discussed in \cite{Drukker:2008zx}.  We present a complete  analysis including the bosonic and fermionic excitations, thus completing some preliminary attempts undertaken in the literature. We find that the systems present some peculiar couplings not seen before in similar situations. 

The rest of the paper is organized as follows. In section~\ref{review} we review the supergravity background and reproduce the leading, classical value of the corresponding D-brane actions.  In section~\ref{D6flucts} we study the fluctuations of the D6 brane in both its bosonic and fermionic sectors and summarize the spectrum of dual operators.
The analysis of the D2 fluctuations and the calculation of the corresponding spectrum of dual operators is carried out in section~\ref{D2flucts}. In section~\ref{comments}, we relate our findings to the structure of supersymmetric multiplets known from the literature, and we conclude in section~\ref{Sec:Conclusions}. We treat a number of more technical and additional aspects in a series of appendices. In particular, in appendix~\ref{cpn} we review the metric representations of $\cpspace^n$ needed in the main text. In appendix~\ref{supergroup} we recall some details of the representation of $OSp(4|2)$.
The harmonic analysis on the coset space $\Toneone$, which we need for the D6 fluctuations, is presented in appendix~\ref{harmonic}.

%% file: sec2.tex
\section{Background configurations}
\label{review}

\subsection{SUGRA background}

We start by reviewing the $\adsfour\times\cpthree$ solution of type-IIA SUGRA, which is the dual of the ABJM theory \cite{Aharony:2008ug}. This solution was described more than three decades ago by Nilsson and Pope \cite{Nilsson:1984bj} and we rely heavily on their presentation. 

Our conventions will be as follows. We work with a Minkowski metric with 
$(- + \ldots +)$ signature. 
The $\adsfour$ and $\cpthree$ coordinates are denoted by the sets of indices $0,1,2,3$ and $4,\ldots,9$, respectively. The corresponding flat indices are underlined. Moreover, we set $\alpha'=1$. 

For our analysis, we shall use the string frame expressions for the background geometry given in \cite{Drukker:2008zx}, but we find it more convenient to work with dimensionless fields. Given the scope of the manuscript we start by considering  the bosonic D$p$-brane action, which, in Minkowski signature, is given by
\begin{equation}
\label{cd:Dp.action}
	S_{\mathrm{D}p}^{B} = - T_p \int d^{p+1} \xi\, \e{-\Phi} \sqrt{- \det(g_{ab}+\cf_{ab})} + T_p \int \e{\cf} \wedge \sum_q C_q~.
\end{equation}
with $\cf_{ab}=B_{ab}+2\pi F_{ab}$, and $T_p=(2\pi)^{-p}$ is the D$p$-brane tension. The metric $g_{ab}$, the 2-form $B_{ab}$ and the RR fields $C_q$ are intended as the pull-backs of the respective 10$d$ background fields. 

The $\adsfour\times\cpthree$ solution is given by \cite{Drukker:2008zx}
\begin{equation}
\label{cd:metric.10D}
	d s^2_{10} = \frac1\beta \left( ds_\adsfour^2 + d\Sigma_3^2 \right)~, \quad 
	\e{2\Phi_0} = \frac{4}{\beta k^2}~,\quad
	F_4 = \frac{3k}{2\beta} \epsilon_\adsfour~,\quad
	F_2 = \frac{k}2 J_3~.
\end{equation}
Here, $d\Sigma_3$ and $J_3$ are the line element and the K\"ahler form of unit-2 $\cpspace^3$, respectively, see appendix~\ref{cpn} for the definitions. $\adsfour$ is of unit radius. The relations to the dual field theory parameters and to the parameters used in \cite{Drukker:2008zx} are
\begin{equation}
\label{bg:beta.lambda}
	\beta = \frac{4k}{R^3} = \left(\pi\sqrt{2\lambda}\right)^{-1}~,\qquad \lambda = \frac{N}k.
\end{equation}
This suggests the following rescalings,
\begin{equation}
\label{cd:rescalings}
	d\hat{s}^2_{10} = \beta ds^2_{10}~,\quad 
	\hat{\Phi} = \Phi-\Phi_0~,\quad
	\hat{\cf}=\beta \cf~,\quad
	\hat{C}_p= \e{\Phi_0}\beta^{\frac{p}2} C_p~.
\end{equation}
Thus, the action \eqref{cd:Dp.action} becomes
\begin{equation}
\label{cd:Dp.action.new}
	S_{\mathrm{D}p}^{B} = - \hat{T}_p \int d^{p+1} \xi\, \e{-\hat{\Phi}} \sqrt{-\det(\hat{g}_{ab}+\hat{\cf}_{ab})} 
		+ \hat{T}_p \int \e{\hat{\cf}} \wedge \sum_q \hat{C}_q~,
\end{equation}
where the D$p$-brane tension $\hat{T}_p$ is now
\begin{equation}
\label{cd:T}
	\hat{T}_p = T_p \e{-\Phi_0} \beta^{-\frac{p+1}2}~.
\end{equation}
In particular,
\begin{equation}
\label{cd:T2.6} 
	\hat{T}_2 = \frac{N}{4\pi \sqrt{2\lambda}} = \frac14\beta N~,\qquad 
	\hat{T}_6 = \frac{N\sqrt{2\lambda}}{(4\pi)^3} = \frac{N}{(4\pi)^3\pi \beta}~.
\end{equation}
The same rescaling procedure can be applied on the fermion action.
Henceforth, we shall omit the hats for simplicity.

Applying \eqref{cd:rescalings} to \eqref{cd:metric.10D}, we find the dimensionless expressions
\begin{equation}
\label{review:10D}
	d s^2_{10} = ds_\adsfour^2 + d\Sigma_3^2~, \quad 
	\Phi =0~,\quad
	F_4 = 3 \epsilon_\adsfour~,\quad
	F_2 = J_3~.
\end{equation}
The dual field strengths are given by\footnote{In our conventions, the K\"ahler form is 
\begin{equation}
\label{review:F2.form}
	J_3 = -\left( e^\flatind{4} \wedge e^\flatind{9} + e^\flatind{5} \wedge e^\flatind{6} + e^\flatind{7} \wedge e^\flatind{8} \right)~. 
\end{equation}}
\begin{equation}
\label{review:dual.F}
	F_6 = \ast F_4= -3\epsilon_{\cpthree}~,\qquad F_8 = \ast F_2 = -\frac12 \epsilon_\adsfour\wedge F_2 \wedge F_2~,
\end{equation}
where $\epsilon_{\cpthree}$ denotes the volume form of the unit-2 $\cpthree$. 

To conclude this review, we recall from \cite{Drukker:2008zx} the explicit  expressions for the metric which we will use in this manuscript\footnote{For $\cpthree$, this is the $m=n=1$ foliation \eqref{cpn:cp3.1}. The $\cpthree$ coordinates take values $\alpha, \vartheta_{1,2} \in (0,\pi)$, $\varphi_{1,2} \in (0,2\pi)$, $\chi \in (0,4\pi)$.}
\begin{align}
\notag
	d s^2_\adsfour &= \cosh^2u\, d s^2_\adstwo + d u^2 + \sinh^2u\, d\phi^2~, \\
\label{review:metrics}
	d \Sigma_3 &= d\alpha^2+ \cos^2\frac{\alpha}{2}(d\vartheta_1^2 +\sin^2\vartheta_1\,d\varphi_1^2)
	+ \sin^2\frac{\alpha}{2}(d\vartheta_2^2 +\sin^2\vartheta_2\,d\varphi_2^2) \\
\notag
	&\quad + \sin^2\frac{\alpha}{2}\cos^2\frac{\alpha}{2}
	(d\chi +\cos\vartheta_1\, d\varphi_1 -\cos\vartheta_2\, d\varphi_2)^2~,
\end{align}
and the $C$-forms
\begin{align}
\notag 
	C_1 &= \frac12 (\cos\alpha-1) d\chi + \cos^2 \frac{\alpha}2 \cos \vartheta_1 d\varphi_1 +  
		\sin^2 \frac{\alpha}2 \cos \vartheta_2 d\varphi_2~,\\ 
\notag 
	C_3 &= \left(\cosh^3 u-1 \right) \epsilon_\adstwo \wedge d \phi~,\\
\notag
	C_5 &= \frac18 (\sin^2\alpha\cos\alpha +2\cos\alpha -2) \sin\vartheta_1 \sin\vartheta_2\, 
	d\vartheta_1\wedge d\varphi_1 \wedge d\vartheta_2 \wedge d\varphi_2 \wedge d\chi~,\\
\label{review:C.forms}
	C_7 &= -\frac16 \left(\cosh^3u-1 \right) \epsilon_\adstwo\wedge d\phi \wedge F_2 \wedge F_2~.
\end{align}
Our conventions for the volume forms are
\begin{align}
\notag
	\epsilon_{(10)} &= \epsilon_\adsfour \wedge \epsilon_{4\cpthree}~,\\
\label{review:volumes}
	\epsilon_\adsfour &= \cosh^2u\, \sinh u\, \epsilon_\adstwo \wedge du\wedge d\phi~,\\
\notag
	\epsilon_{\cpthree} &= \frac18 \sin^3\alpha \sin\vartheta_1 \sin\vartheta_2\, d\alpha \wedge
		d\vartheta_1 \wedge d\varphi_1 \wedge d\vartheta_2 \wedge d\varphi_2 \wedge d\chi~.
\end{align}

It is known that there are two inequivalent $\adsfour\times\cpthree$ solutions of IIA SUGRA, one $\mathcal{N}=6$ supersymmetric, the other one without supersymmetries \cite{Nilsson:1984bj}. We are, of course, interested in the $\mathcal{N}=6$ solution. The difference between the two solutions lies in a relative sign of $F_2$ and $F_4$, and one is well advised, in view of diverse conventions, to check the supersymmetry of the above configuration. For doing so, we use the supersymmetry transformations given in \cite{Martucci:2005rb}, because we will rely on that paper for the construction of the fermion action. The supersymmetry transformation of the gravitino and dilatino are 
\begin{equation}
\label{review:SUSY}
	\delta_\epsilon \psi_m = D_m \epsilon~, \qquad \delta_\epsilon \lambda = \Delta \epsilon~,
\end{equation} 
where (dropping terms that vanish in our case)
\begin{align}
\label{review:SUSY1}
	D_m &= \nabla_m - \frac18 \left( \frac12 F_{np} \Gamma^{np} \Gamma_{(10)} +\frac1{4!} F_{npqr} \Gamma^{npqr} \right) \Gamma_m~,\\
\label{review:SUSY2}
	\Delta &= \frac18 \left( \frac32 F_{np} \Gamma^{np} \Gamma_{(10)} - \frac1{4!} F_{npqr} \Gamma^{npqr} \right)~.
\end{align}
The 10$d$ chirality matrix is defined by $\Gamma_{(10)}= \Gamma^\flatind{0\cdots9}$. To check whether \eqref{review:10D} is supersymmetric, one first considers the dilatino variation in \eqref{review:SUSY}. Defining 
\begin{equation}
\label{review:Q.def}
	Q= \frac12 F_{mn} \Gamma^{mn} \Gamma^\flatind{456789} = \Gamma^\flatind{5678} + \Gamma^\flatind{4569}+ \Gamma^\flatind{4789}~, 
\end{equation}
and using \eqref{review:10D}, \eqref{review:SUSY2} can be written as
\begin{equation}
\label{review:Delta}
	\Delta = \frac38 \Gamma^\flatind{0123} (Q-1)~.
\end{equation}
Moreover, it follows from \eqref{review:Q.def} that $Q$ satisfies 
\begin{equation}
\label{review:Q.prop}
	(Q+3)(Q-1) =0~,
\end{equation}
and has the eigenvalues $(-3,-3,1,1,1,1,1,1)$. The degeneracies follow from $\tr Q=0$. There are, therefore, six $\cpthree$ spinors that solve 
\begin{equation}
\label{review:Q.epsilon}
	Q\epsilon = \epsilon~.
\end{equation}
Comparing with \cite{Nilsson:1984bj} we find that this is indeed the $\mathcal{N}=6$ solution. We also recall from \cite{Nilsson:1984bj} that the $\adsfour$ components of \eqref{review:SUSY1} yield four $\adsfour$ Killing spinors, and that by virtue of  \eqref{review:Q.epsilon} the integrability condition for the $\cpthree$ components of \eqref{review:SUSY1} is satisfied.

\subsection{D6 and D2-branes}
\label{review:BPS16}


The D6-brane purportedly dual to the 1/6 BPS totally antisymmetric Wilson loop wraps $\adstwo\subset\adsfour$ at the point $u=0$ and $\Toneone\subset\cpthree$ at constant $\alpha$. The latter is a squashed $T^{1,1}$ space \cite{Benincasa:2011zu}. The internal gauge field $\cf$ has electric flux only in the $\adstwo$ factor, $\cf=E\epsilon_\adstwo$, where $E$ is conjugate to the fundamental string charge $p$. Because the latter is fixed, the potential that yields the Wilson loop expectation value is the  Legendre transform of the D6-brane action \cite{Drukker:2008zx}. It is straightforward to obtain\footnote{The renormalized volume of the unit $\adstwo$ is $V_\adstwo = -2\pi$ \cite{Faraggi:2011ge}.} 
\begin{equation}
\label{review:D6.action.2}
	S_{\text{WL}} = S_\Dsix^B -\frac{1}{\beta} p E = \frac{N}{4\beta} \left[ \sin^3 \alpha \sqrt{1-E^2} 
	- E \left(\sin^2\alpha\cos\alpha + 2\cos\alpha -2\right) \right]-\frac{1}{\beta} p E~.
\end{equation}
The equation of motion for $\alpha$ fixes 
\begin{equation}
\label{review:E}
	E = - \cos\alpha~,
\end{equation}
and that for $E$ yields 
\begin{equation}
	p= \beta \frac{\delta S_\Dsix^B}{\delta E} = \frac{N}2 (1-\cos\alpha)~.
\end{equation}
The fact that $p$ ranges from  0 to $N$ is a signature of the antisymmetric representation. This evidence for the anti-symmetric representation is a typical phenomenon in many brane configurations originally understood in the case of the giant gravitons \cite{McGreevy:2000cw,Hashimoto:2000zp}. Finally, the expectation value of the Wilson loop is found as
\begin{equation}
	S_{\text{WL}} = \frac{p(N-p)}{\beta N}.
\end{equation}
Note the symmetry under $p\leftrightarrow N-p$. It was shown in \cite{Drukker:2008zx} that this D6-brane is $1/6$-BPS.


The D2-brane dual to the 1/6 BPS symmetric Wilson loop wraps $\adstwo\subset\adsfour$ at the point $u=0$ and the circle $S^1\subset \cpthree$ along $\chi$. Again, $\cf=E\epsilon_\adstwo$. With this configuration, the Wilson loop potential is 
\begin{equation}
\label{review:D2.action.2}
	S_{WL} = S_\Dtwo^B -\frac{1}{\beta} p E = \beta N \pi^2 \left[ \sin \alpha \sqrt{1-E^2} 
	- E \left(\cos\alpha -1 \right) \right]-\frac{1}{\beta} p E~.
\end{equation}
The field equation for $\alpha$ yields again \eqref{review:E}, while the equation for $E$ yields  
\begin{equation}
	p= \beta \frac{\delta S_\Dtwo^B}{\delta E} = \beta^2 N \pi^2 = \frac12 k~,
\end{equation}
corresponding to $k/2$ fundamental strings. Finally, the Wilson loop expectation is
\begin{equation}
	S_{WL}= \frac{k}2 \sqrt{2\lambda} \pi~.
\end{equation}
It was shown in \cite{Drukker:2008zx} that a single D2-brane is $1/3$-BPS. Smearing on $\cpspace^1$ reduces supersymmetry to $1/6$-BPS. There are outstanding questions as to in which precise higher rank representation each of the classical solutions discussed here and their generalizations reside. We leave a precise study of these questions to a separate publication. Let us simply note that other possible classical configurations do not seem to fit nicely with their $AdS_5\times S^5$ counter-part. For example, the symmetric representation in that case corresponds to a D3 brane discussed in \cite{Drukker:2005kx} whose spectrum of quantum excitations was presented in \cite{faraggi:2011bb}.  This D3 branes wraps $AdS_2\times S^2 \subset AdS_5$ and the value of its electric flux can be arbitrarily large. We have verified that the analogous D2 configuration wrapping the $AdS_2\times S^1\subset AdS_4$ does not seem to have the expected properties.

The beautiful construction of the 1/2 BPS Wilson loop on the field theory side \cite{Drukker:2009hy} and some of its generalizations discussed in \cite{Cardinali:2012ru} are still largely unexplored on the holographic side; the gap is particularly glaring in the case of higher rank representations. Let us advance a few observations we have briefly explored in this regard.  On grounds of the supergroup symmetries, one expects that the 1/2 BPS D6 configuration should wrap $\mathbb{CP}^2\subset \mathbb{CP}^3$ as to have $U(3)$ symmetry realized in its worldvolume. Correspondingly, there are potential D2 configurations that wrap a circle transverse to $\mathbb{CP}^2\subset \mathbb{CP}^3$ and therefore, contain the action of $U(3)$ in the flucutations transverse to the worldvolume.    A very preliminary exploration of these possibilities also yields puzzling results and we will report on these configurations separately.

%% file: sec3.tex
\section{D6-brane fluctuations}
\label{D6flucts}

In this section, we consider the bosonic and fermionic fluctuations of the $1/6$-BPS D6-branes. 
The notation in this section will be as follows. The 10$d$ curved coordinates are denoted by Latin indices from the middle of the alphabet, $m,n=0,\ldots,9$. Latin indices from the beginning of the alphabet denote generic $\Dsix$-brane coordinates, $a,b=0,1,5,6,7,8,9$.
When the worldvolume is split into $\adstwo \times \mathcal{M}_5$, $\alpha,\beta=0,1$ are used for the $\adstwo$ part, while Greek indices from the middle of the alphabet, $\mu,\nu=5,\ldots,9$, are reserved for the factor $\mathcal{M}_5 \subset \cpthree$. Latin indices $i,j=2,3,4$ denote the normal directions. Flat indices are underlined. 

\subsection{Bosonic fluctuations}

For the bosonic fluctuations, we start with the action \eqref{cd:Dp.action.new}. We follow 
the procedure described in detail in \cite{Faraggi:2011ge}, which relies on the geometry of embedded manifolds and renders all expressions manifestly covariant. We refer the reader to Sec.~3 and Appendix~B of that paper for the relevant formulae. Following this strategy, the fluctuations of the $\Dsix$-brane worldvolume are parameterized by three scalars $\chi^\flatind{i}$ corresponding to the three normal directions. They consist of a doublet $(i=2,3)$ characterizing the normals of $\adstwo \subset \adsfour$ and a singlet $(i=4)$ for the normal within $\cpthree$. The worldvolume displacement is described by a geodesic map,
\begin{equation}
\label{D6flucts:geod.map}
	x^m \to (\exp_x y)^m~,\qquad y^m = N^m_\flatind{i} \chi^\flatind{i}~.
\end{equation}
In addition, there are the fluctuations of the 2-form gauge field, 
\begin{equation}
\label{D6flucts:F}
	\cf_{ab} \to \cf_{ab} +f_{ab}~, \quad f = da~.
\end{equation}
Defining $M_{ab}=g_{ab}+\cf_{ab}$, we have to second order [cf.\ (3.10) of \cite{Faraggi:2011ge}]
\begin{equation}
\label{D6flucts:delta.M}
	\delta M_{ab} = -2H_{\flatind{i}ab} \chi^\flatind{i} + f_{ab}
	+ \nabla_a \chi^\flatind{i} \,\nabla_b \chi^\flatind{j} \,\delta_\flatind{ij}
	+  \left( H_{\flatind{i}a}{}^c H_{\flatind{j}bc} -R_{mpnq} x^m_a x^n_b N^p_\flatind{i} N^q_\flatind{j} \right)
	\chi^\flatind{i} \chi^\flatind{j}~.
\end{equation}
Here, $H^\flatind{i}_{ab}$ is the extrinsic curvature (second fundamental form) of the embedding. 
The expansion up to second order of the Born-Infeld (BI) term may be obtained from the general formula
\begin{equation}
\label{D6flucts:M.expand}
	\sqrt{-\det M} \to \sqrt{-\det M}\left[ 1+\frac12 \tr X + \frac18 (\tr X)^2 - \frac14 \tr X^2 \right]~,
\end{equation}
where $X= M^{-1}\delta M$. This yields 
\begin{align}
\label{D6:BI}
	\sqrt{-\det M_{ab}} &\to \sqrt{-\det g_{ab}} \sin\alpha \Bigg\{ 1 + 3 \cot\alpha\, \chi^\flatind{4}
	-\frac{\cos\alpha}{\sin^2\alpha} \left(\frac12 \epsilon^{\alpha\beta} f_{\alpha\beta}\right)\\
\notag
	&\quad +\frac1{2\sin^2\alpha} \nabla^\alpha \chi^\flatind{i} \nabla_\alpha \chi_\flatind{i}  
	+\frac12 \nabla^\mu \chi^\flatind{i} \nabla_\mu \chi_\flatind{i} \\
\notag 
	&\quad +\frac1{\sin^2\alpha} \left[(\chi^\flatind{2})^2+ (\chi^\flatind{3})^2\right] 
	+ \left(\frac3{\sin^2\alpha}-\frac92\right) (\chi^\flatind{4})^2\\
\notag
	&\quad +\frac1{4\sin^4\alpha} f_{\alpha\beta}f^{\alpha\beta} +\frac14 f_{\mu\nu}f^{\mu\nu} 
	+\frac1{2\sin^2\alpha} f^{\alpha\mu}f_{\alpha\mu}
	- \frac{3\cos^2\alpha}{\sin^3\alpha} \chi^\flatind{4} \left(\frac12 \epsilon^{\alpha\beta} f_{\alpha\beta}\right) \Bigg\}~.
\end{align}
Here we have used
\begin{equation}
\label{D6:H4}
	H_\flatind{i}{}_{\alpha\beta}=0~, \qquad
	H_\flatind{4}{}^\mu{}_\mu = -3 \cot\alpha~, \qquad 
	H_\flatind{4}{}^{\mu\nu}H_\flatind{4}{}_{\mu\nu} = 3 \cot^2 \alpha +1~,
\end{equation} 
and the fact that $\cpthree$ is Einstein, $R^{4\cpthree}_{mn} = \frac{2\cdot 3+2}4 g^{4\cpthree}_{mn} = 2 g^{4\cpthree}_{mn}$.

The Wess-Zumino (WZ) terms are obtained taking into account the expansion of the form fields and the tangent vectors for the pull-back, cf.\ (3.3) and (3.4) of \cite{Faraggi:2011ge}. The $C_7$ WZ term gives
\begin{equation}
\label{D6:WZ7}
	P[C_7] \to d^7 \xi\sqrt{-\det g_{ab}} \frac12 e^\mu_\flatind{9} 
	\left( \chi^\flatind{2} \nabla_\mu \chi^\flatind{3} - \chi^\flatind{3} \nabla_\mu \chi^\flatind{2} \right)~,
\end{equation}
where the indices $\flatind{2}$ and $\flatind{3}$ denote the normals in the $u$-- and $\phi$--directions, respectively.
This contribution is somewhat unexpected, because both $C_7$ and its first $u$--derivative vanish for $u=0$. However, one must carefully consider the small--$u$ behaviour, because the normal component $N^\phi_\flatind{3}$ goes like $1/u$. This leads to the finite result \eqref{D6:WZ7}, which is absent in previous discussions of similar classical configurations. 

The $C_5$ WZ term leads to 
\begin{align}
\label{D6:WZ5}
	 \cf\wedge P[C_5] &\to d^7 \xi\sqrt{-\det g_{ab}} \Bigg\{ -\cos\alpha\, C(\alpha)  
	+3 \cos \alpha\, \chi^\flatind{4} -C(\alpha) \left(\frac12 \epsilon^{\alpha\beta} f_{\alpha\beta}\right) \\
\notag 
	&\quad +3 \chi^\flatind{4} \left(\frac12\epsilon^{\alpha\beta} f_{\alpha\beta} \right) 
 	+\frac{9 \cos^2\alpha}{2\sin\alpha} (\chi^\flatind{4})^2 \Bigg\}~,
\end{align}
where 
\begin{equation}
\label{D6:Cdef}
	C(\alpha) = \sin^{-3}\alpha \left( \sin^2\alpha \cos\alpha + 2 \cos\alpha -2 \right)~. 
\end{equation}

The $C_3$ WZ term vanishes, and the $C_1$ WZ term gives a contribution, which is found easily after an integration by parts
\begin{equation}
\label{D6:WZ11}
	\frac16 \cf^3 \wedge P[C_1] = \frac12 \cf \wedge f \wedge f \wedge P[C_1] 
	\to - \frac12 \cos\alpha\, \epsilon_\adstwo \wedge a \wedge f \wedge P[F_2]~,
\end{equation}
where $f=da$. This form has the advantage of being independent of any exact terms in $C_1$. Using \eqref{review:F2.form}, one finds
\begin{equation}
\label{D6:WZ1}
	\frac16 \cf^3 \wedge P[C_1] \to d^7 \xi\sqrt{-\det g_{ab}}\, \frac12 \cos\alpha\, \calE^{\mu\nu\rho} a_\mu \partial_\nu a_\rho~,
\end{equation}
where $\calE^{\mu\nu\rho}$ is the totally antisymmetric tensor known as the Betti 3-form \cite{Witten:1983ux,Ceresole:1999ht},
\begin{equation}
\label{D6:eps3}
	\frac16\calE_{\mu \nu \rho}\, d\xi^\mu \wedge d\xi^\nu \wedge d\xi^\rho 
	= e^\flatind{4} \wedge \left( e^\flatind{5} \wedge e^\flatind{6} + e^\flatind{7} \wedge  e^\flatind{8} \right)~.
\end{equation}

Finally, we sum the contributions \eqref{D6:BI}, \eqref{D6:WZ7}, \eqref{D6:WZ5} and \eqref{D6:WZ1}, drop total derivatives
and express the resulting quadratic action in terms of the open string metric, which rescales the $\adstwo$ part to the radius $\sin\alpha$, 
\begin{equation}
\label{D6:open.string.metric}
	d\tilde{s}^2 = \sin^2\alpha\, g_{\alpha\beta} d\xi^\alpha d\xi^\beta + g_{\mu\nu} d\xi^\mu d\xi^\nu~. 
\end{equation} 
This yields 
\begin{align}
\label{D6:action2}
	S_\Dsix^{B,2} &= -\frac{T_6}{\sin\alpha} \int d^7\xi\, \sqrt{-\det \tilde{g}_{ab}} \Bigg\{ 
	\frac12 \tn^a \chi^\flatind{i} \tn_a \chi^\flatind{i} 
	+\frac1{\sin^2\alpha} \left[ (\chi^\flatind{2})^2 + (\chi^\flatind{3})^2 \right]
	-\frac{3}{2\sin^2\alpha} (\chi^\flatind{4})^2 \\
\notag
	&\quad +\frac1{\sin\alpha} e^\nu_\flatind{9} \chi^\flatind{3} \nabla_\nu \chi^\flatind{2} + \frac14 \tilde{f}^{ab} \tilde{f}_{ab} 
	- \frac{3}{\sin\alpha} \chi^\flatind{4}\left(\frac12\tilde{\epsilon}^{\alpha\beta} f_{\alpha\beta} \right) 
	- \frac12 \cot\alpha\, \calE^{\mu\nu\rho} a_\mu \partial_\nu a_\rho \Bigg\}~,
\end{align}
which is our final result for the bosonic action of the 1/6 BPS $\Dsix$-brane. Note that our result completes a preliminary discussion of the quadratic excitations presented in \cite{Benincasa:2011zu}.


\subsection{Fermionic fluctuations}

For the fermionic fluctuations, our starting point is Eq.~(17) of \cite{Martucci:2005rb},
\begin{equation}
\label{D6flucts:action.F}
	S_\Dsix^F = \frac{T_6}2 \int d^7\xi \e{-\Phi} \sqrt{-\det M_{ab}}\, \bar{\theta} \left(1-\Gamma_\Dsix\right)
	\left[ (\tilde{M}^{-1})^{ab}\Gamma_b D_a -\Delta \right] \theta~,
\end{equation}
where $\theta$ is a 32-component, 10$d$ Majorana spinor, $\bar{\theta}=i\theta^\dagger \Gamma^\flatind{0}$, $\tilde{M}_{ab}=g_{ab}+\Gamma_{(10)} \cf_{ab}$, $D_a=\partial_a X^m D_m$, $D_m$ and $\Delta$ were defined in \eqref{review:SUSY1} and \eqref{review:SUSY2}, respectively, and $\Gamma_\Dsix$ is 
\begin{align}
\notag
	\Gamma_\Dsix &= \frac{\sqrt{-\det g_{ab}}}{\sqrt{-\det (g_{ab}+\cf_{ab})}} \left(-\Gamma^\flatind{0156789}\right)
	\sum_q \frac{ (-\Gamma_{(10)})^q}{q!2^q} \Gamma^{b_1 \ldots b_{2q}} \cf_{b_1b_2} \ldots \cf_{b_{2q-1}b_{2q}}\\
\label{D6flucts:Gamma.D6} 
	&= \frac1{\sin\alpha} \left(-\Gamma^\flatind{0156789}\right) \left( 1 +\cos\alpha\, \Gamma^\flatind{01} \Gamma_{(10)} \right)~.
\end{align}
The pullback of the covariant derivative on spinors is given by \cite{Faraggi:2011ge}
\begin{equation}
\label{D6flucts:D.pullback}
 	\partial_a X^m \nabla_m = \nabla_a -\frac12 H_{\flatind{i}ab}\Gamma^b\Gamma^\flatind{i} +\frac14 A_{\flatind{ij}a} \Gamma^\flatind{ij}~.
\end{equation}
The combinations we need are 
\begin{equation}
\label{D6:D6.ferm.pullbacks}
	\partial_\alpha X^m \nabla_m = \nabla_\alpha~, \quad 
	\Gamma^\mu \partial_\mu X^m \nabla_m = \Gamma^\mu \nabla_\mu +\frac32 \cot\alpha\, \Gamma^\flatind{4}~.
\end{equation}
Direct evaluation of the operator in squared brackets in \eqref{D6flucts:action.F} yields
\begin{align}
\label{D6:operator.explicit}
	(\tilde{M}^{-1})^{ab}\Gamma_b D_a -\Delta &= 
	\frac1{\sin^2\alpha}\left( 1-\cos\alpha\, \Gamma^\flatind{01} \Gamma_{(10)}\right) \Gamma^\alpha \nabla_\alpha
	+\Gamma^\mu \nabla_\mu +\frac32 \cot\alpha\, \Gamma^\flatind{4}\\
\notag &\quad
	+\frac1{4\sin^2\alpha}\left( 1-\cos\alpha\, \Gamma^\flatind{01} \Gamma_{(10)}\right) 
	\left[-\left(\Gamma^\flatind{49} +\Gamma^\flatind{56} +\Gamma^\flatind{78}\right) \Gamma_{(10)}+3 \Gamma^\flatind{0123}\right] \\
\notag &\quad
	+\frac14 \left(\Gamma^\flatind{56} +\Gamma^\flatind{78}\right) \Gamma_{(10)} -\frac32 \Gamma^\flatind{0123}~.
\end{align}

To proceed, we fix the $\kappa$-symmetry by imposing $\theta$ to be chiral. What matters here is that only terms in \eqref{D6flucts:action.F} with an odd number of $\Gamma$-matrices survive the chiral projection. In fact, the chirality is irrelevant.
Hence, we find 
\begin{multline}
\label{D6:chiral.projection}
	\bar{\theta} \left(1-\Gamma_\Dsix\right) \left[ (\tilde{M}^{-1})^{ab}\Gamma_b D_a -\Delta \right] \theta =\\
	\bar{\theta} \e{R\Gamma^\flatind{01}\Gamma_{(10)}} \left[ \tilde{\Gamma}^a \tn_a 
	-\frac14 \cot\alpha \left(\Gamma^\flatind{569}+ \Gamma^\flatind{789}\right) 
	+\frac1{4\sin\alpha} \Gamma^\flatind{239} \left( 1- 3\Gamma^\flatind{5678} \right) \right]
	\e{R\Gamma^\flatind{01}\Gamma_{(10)}} \theta~,
\end{multline}
where the spinor rotation parameter $R$ is determined by $\sinh 2R=-\cot\alpha$. In what follows, we simply work with the rotated spinor, $\e{R\Gamma^\flatind{01}\Gamma_{(10)}} \theta \to \theta$. The Dirac operator in \eqref{D6:chiral.projection} is the one corresponding to the open string metric \eqref{D6:open.string.metric}. 

To proceed, it is necessary to decompose the $32\times 32$ $\Gamma$-matrices into a 7$d$ representation. We shall use 
\begin{align}
\notag
	\Gamma^a &= \gamma^a \otimes \mathbb{I}_2  \otimes \sigma_1 ~, &&(a=0,1,5,6,7,8,9)\\
\label{D6:decomp}
	\Gamma^\flatind{i} &= \mathbb{I}_8 \otimes \tau_{i-1} \otimes \sigma_2 ~, &&(i=2,3,4)~,
\end{align}
where $\gamma^a$, $\tau_i$ and $\sigma_i$ denote 7$d$ Minkowski Gamma matrices and two copies of Pauli matrices, respectively. The representation \eqref{D6:decomp} is chiral with, 
\begin{equation}
\label{D6:chiral.rep}
	\Gamma_{(10)} = - \gamma^\flatind{0156789} \otimes \mathbb{I}_2 \otimes \sigma_3
	= \pm \mathbb{I}_8 \otimes \mathbb{I}_2 \otimes \sigma_3~,
\end{equation}
where the sign depends on the representation of the 7$d$ gamma matrices. Hence, a $10d$ chiral spinor (16 components) decomposes into a doublet of 7$d$ spinors, and the matrices $\tau_i$ act on the doublet. 

The Majorana condition on $\theta$ translates into a symplectic Majorana condition on the 7$d$ spinor doublet. To see this, decompose the  the Majorana intertwiner \cite{VanProeyen:1999ni} into
\begin{equation}
\label{D6:intertwiner}
	B_{+(9,1)} = B_{+(6,1)} \otimes B_{-(3,0)} \otimes \mathbb{I}_2~.
\end{equation}

Finally, after applying the decomposition \eqref{D6:decomp} to \eqref{D6:chiral.projection} and substituting the result into \eqref{D6flucts:action.F}, we obtain the fermionic action
\begin{equation}
\label{D6:ferm.action2}
	S_\Dsix^F = \frac{T_6}{2\sin\alpha} \int d^7\xi \sqrt{-\det \tilde{g}_{ab}}\, \bar{\theta}_\pm
	\Bigg[ \tilde{\gamma}^a \tn_a -\frac14 \cot\alpha \left(\gamma^\flatind{569}+ \gamma^\flatind{789}\right) 
	\pm \frac{i}{4\sin\alpha} \gamma^\flatind{9} \left( 1- 3\gamma^\flatind{5678} \right) \Bigg] \theta_\pm~.
\end{equation}
There is an implicit sum over the spinor doublet index ($\pm$), and the sign of the last term in the brackets agrees with the doublet index. 

We conclude this section by writing Eq. \eqref{D6:ferm.action2} in a $2+5$ form, which is useful for the calculation of the spectrum. We shall use the decomposition
\begin{equation}
\label{D6:decomp2}
	\gamma^\alpha = \gamma^\alpha \otimes \mathbb{I}_4~,\qquad 
	\gamma^\mu = \gamma^\flatind{01} \otimes \gamma^\mu~,
\end{equation}
where the matrices $\gamma^\alpha$ and $\gamma^\mu$ on the right hand sides are intended as 2$d$ and 5$d$ gamma matrices, respectively. 
Hence, we can rewrite \eqref{D6:ferm.action2} as
\begin{equation}
\label{D6:ferm.action3}
	S_\Dsix^F = \frac{T_6}{2\sin\alpha} \int d^7\xi \sqrt{-\det \tilde{g}_{ab}}\, \bar{\theta}_\pm\,
	\left( \tilde{\gamma}^\alpha \tn_\alpha \otimes \mathbb{I}_4 
	+ \gamma^\flatind{01} \otimes \mathcal{D}_\pm \right) \theta_\pm~,
\end{equation}
where the differential operators $\mathcal{D}_\pm$ acting on the $\Toneone$ part are
\begin{equation}
\label{D6:ferm.D.ops}
	\mathcal{D}_\pm = \tilde{\gamma}^\mu \tn_\mu 
	-\frac14 \cot\alpha \left(\gamma^\flatind{569}+ \gamma^\flatind{789}\right) 
	\pm \frac{i}{4\sin\alpha} \gamma^\flatind{9} \left( 1- 3\gamma^\flatind{5678} \right)~.
\end{equation}

\subsection{Field equations}
\label{D6.field.equations}

For completeness, we list here the field equations deriving from the actions \eqref{D6:action2} and \eqref{D6:ferm.action3}. The doublet of scalars $\chi^\flatind{i}$, $i=2,3$, satisfy
\begin{align}
\label{eom:scal.1}
	\left( -\tn_a \tn^a +\frac{2}{\sin^2\alpha} \right) \chi^\flatind{2} 
	-\frac1{\sin\alpha} e^\mu_\flatind{9} \tn_\mu \chi^\flatind{3} &=0~,\\ 
\label{eom:scal.2}
	\left( -\tn_a \tn^a +\frac{2}{\sin^2\alpha} \right) \chi^\flatind{3} 
	+\frac1{\sin\alpha} e^\mu_\flatind{9} \tn_\mu \chi^\flatind{2} &=0~.
\end{align} 
Introducing $\chi^\pm = \chi^\flatind{2}\pm i \chi^\flatind{3}$, \eqref{eom:scal.1} and  \eqref{eom:scal.2} become
\begin{equation}
\label{eom:scal}
	\left( -\tn_a \tn^a +\frac{2}{\sin^2\alpha} \pm \frac{i}{\sin\alpha} e^\mu_\flatind{9} \tn_\mu \right) \chi^\pm =0~. 
\end{equation}
It is worth noting that this is a generalization of what would traditionally be a couple of massive fields describing the embedding of $AdS_2 \subset AdS_4$. Namely, in the absence of the last term above, one has two scalar fields with $m^2=2$ just as in the case \cite{Sakaguchi:2010dg}. Similarly for the embedding of supersymmetric branes in $AdS_5\times S^5$, one gets three $m^2=2$ modes from  $AdS_2\subset AdS_5$ for the D3 and D5 respectively \cite{faraggi:2011bb,Faraggi:2011ge}. It is easy to track this term to the $C_7$  contribution from  the WZ part of the action (see Eq. \ref{D6:WZ7}); we will see that there is a corresponding $C_3$ contribution to the D2 fluctuations, thus leading to a sort of universality.

The scalar $\chi^\flatind{4}$ couples to the $\adstwo$-components $a_\alpha$ of the vector field. Their field equations are given by
\begin{align}
\label{eom:scal.vec.1}
	\left( \tn_a \tn^a +\frac{3}{\sin^2\alpha} \right) \chi^\flatind{4} +\frac3{\sin\alpha} f &= 0~, \\
\label{eom:scal.vec.2}
	 \tn_a (\tn^a a^\alpha -\tn^\alpha a^a) + \frac3{\sin\alpha} \tilde{\epsilon}^{\alpha\beta} \partial_\beta \chi^\flatind{4} &=0~,
\end{align}
where $f$ stands for $f=\frac12 \tilde{\epsilon}^{\alpha\beta} f_{\alpha\beta}$. We adopt the Lorentz gauge, $\tn_a a^a =0$. The remaining gauge freedom can be used to further impose $\tn_\alpha a^\alpha = \tn_\mu a^\mu=0$ on-shell. Acting with $\tn^\gamma \tilde{\epsilon}_{\gamma\alpha}$ on \eqref{eom:scal.vec.2}, one obtains
\begin{equation}
\label{eom:scal.vec.3}
	\tn_a \tn^a f + \frac{3}{\sin\alpha} \tn_\alpha \tn^\alpha \chi^\flatind{4} =0~.
\end{equation}
Hence, we can write \eqref{eom:scal.vec.1} and \eqref{eom:scal.vec.3} in the matrix form
\begin{equation}
\label{eom:scal.vec.matrix}
	\begin{pmatrix}
	\tn_\alpha \tn^\alpha +\tn_\mu \tn^\mu +\frac{3}{\sin^2\alpha} & \frac3{\sin\alpha} \\
	\frac3{\sin\alpha} \tn_\alpha\tn^\alpha & \tn_\alpha \tn^\alpha +\tn_\mu \tn^\mu 
	\end{pmatrix}
	\begin{pmatrix} \chi^\flatind{4} \\ f \end{pmatrix} =0~.
\end{equation}

The vector components $a^\mu$ satisfy, in Lorentz gauge, 
\begin{equation}
\label{eom:vec}
	-\left( \tn_\alpha \tn^\alpha + \tn_\nu \tn^\nu \right) a^\mu + R^\mu{}_\nu a^\nu 
	-\cot\alpha\, \calE^{\mu\nu\rho} \partial_\nu a_\rho =0~.
\end{equation}

The field equations for the spinors are simply
\begin{equation}
\label{eom:spin}
	\left( \tilde{\gamma}^\alpha \tn_\alpha \otimes \mathbb{I}_4 
	+ \gamma^\flatind{01} \otimes \mathcal{D}_\pm \right) \theta_\pm~=0,
\end{equation}
where $\mathcal{D}_\pm$ is defined by \eqref{D6:ferm.D.ops}.

\subsection{Spectrum of D6-brane fluctuations}
\label{spec}

In this section, we calculate the spectrum of fluctuations of the $\Dsix$-brane and obtain the conformal dimensions of the dual operators. The bosonic fluctuations were considered in \cite{Benincasa:2011zu}, but the result is partially incorrect because of missing terms in the quadratic action. 
To obtain the spectrum, the equations of motion listed in subsection~\ref{D6.field.equations} must be solved. This requires to construct the (generalized) harmonics on the $\Toneone$ factor of the D6 world volume, which we defer to appendix~\ref{harmonic} due to its rather technical nature. 

We start with the doublet of scalars, $\chi^\flatind{i}$, ($i=2,3$). The field equation for the combinations $\chi^\pm=\chi^\flatind{2}\pm i\chi^\flatind{3}$ is given by \eqref{eom:scal}. Substituting \eqref{harmonic:scal.modified.laplace} and \eqref{harmonic:H.rewrite}, it becomes a field equation on $\adstwo$,
\begin{equation}
\label{spectrum:eom.scal}
	\left( \tn^\alpha \tn_\alpha - m_\pm^2 \right) \chi^\flatind{\pm} =0~,
\end{equation}
where 
\begin{equation}
\label{spectrum:eom.scal.mass}
	m_\pm^2 = \frac{\cll +1\pm r}{\sin^2\alpha}~.
\end{equation}
Because the radius of $\adstwo$ in the open string metric is $\sin\alpha$, the standard relation between $m^2$ and the conformal dimension of the dual operator yields   
\begin{equation}
\label{spec:dim.scal}
	\Delta^{(\pm)} = \frac12 +\sqrt{\frac54+\cll\pm r}~.
\end{equation}
We recall the definition \eqref{harmonic:H.rewrite} of $\cll$,
\begin{equation}
\label{spec:cll.def}
	\cll = \sin^2\frac{\alpha}2 (2j+1)^2 +\cos^2\frac{\alpha}2 (2l+1)^2~.
\end{equation}
As explained in appendix~\ref{harmonic:spectrum}, $j$, $l$ are either both integer or half-integer, and $|r|\leq \bar{l}$, where 
\begin{equation}
\label{spec:lbar.def}
	\bar{l} = 2 \min(j,l)~.
\end{equation}

The field equations of the scalar $\chi^\flatind{4}$ and the $\adstwo$-components of the vector field are given by \eqref{eom:scal.vec.matrix}. Substituting the eigenvalues of the scalar Laplacian on $\Toneone$ \eqref{harmonic:scalar.laplace},
one obtains 
\begin{equation}
\label{spec:eom.scal.vec.matrix}
	\begin{pmatrix}
	-\tn_\alpha \tn^\alpha + \frac{\cll-4}{\sin^2\alpha} & -\frac3{\sin\alpha} \\
	-\frac3{\sin\alpha} \tn_\alpha\tn^\alpha & -\tn_\alpha \tn^\alpha + \frac{\cll-1}{\sin^2\alpha}
	\end{pmatrix}
	\begin{pmatrix} \chi^\flatind{4} \\ f \end{pmatrix} =0~.
\end{equation}
The characteristic polynomial of this matrix is equivalent to the product of two massive Klein-Gordon equations on $\adstwo$, with two mass values. To these correspond the following conformal dimensions of the two dual operators,
\begin{equation}
\label{spec:dim.scal.vec}
	\Delta^{(\flatind{4})} \in \left\{ \sqrt{\cll}+2; \sqrt{\cll}-1\right\}~.
\end{equation}
The eigenvalues are $(\bar{l}+1)$-fold degenerate, because they are independent of $r$.
From the second value one must exclude the case $j=l=0$ ($\cll=1$), because the corresponding bulk mode is not dynamical \cite{Benincasa:2011zu,Faraggi:2011ge}. (It is the gauge mode that allows to impose $\tn_\alpha a^\alpha =\tn_\mu a^\mu=0$, which is more restrictive than the Lorentz gauge $\tn_a a^a=0$.) 

Consider the $\Toneone$ components of the vector field. Their field equations are given by \eqref{eom:vec}, which becomes a massive Klein-Gordon equation on $\adstwo$ of the form \eqref{spectrum:eom.scal} (the $\Toneone$ vector is an $\adstwo$ scalar) once the results of the harmonic analysis on $\Toneone$ have been used. The mass-square is simply given by the eigenvalues of the modified vector Laplacian, which are listed in appendix~\ref{app:tables}. The conformal dimension of the dual operator then follows from the standard formula. We list the results in Tables~\ref{spec:table.dims.gen} and \ref{spec:table.dims.sp} for the generic case $j\neq l$ and the special case $j=l$, respectively.

The conformal dimensions of the operators dual to the spinor fields are found from the spinor field equation \eqref{eom:spin}. After using the results of the harmonic analysis, one may consider
\begin{equation}
\label{spec:eom.spin}
	\left( \tilde{\gamma}^\alpha \tn_\alpha + \lambda \gamma^\flatind{01} \right) \vartheta \otimes \theta_\lambda~,
\end{equation}
where $\lambda=ich$ represents the eigenvalue of $\mathcal{D}_\pm$ corresponding to the eigenvector $\theta_\lambda$, which is a $\Toneone$ spinor, while $\vartheta$ is a spinor on $\adstwo$. Denoting by $\vartheta_\mu$ ($\mu\geq 0$) a solution of the $\adstwo$ Dirac equation
\begin{equation}
\label{spec:eom.spin.ads}
	\left( \tilde{\gamma}^\alpha \tn_\alpha  -\mu \right) \vartheta_\mu =0~,
\end{equation}
and using $\gamma^\flatind{01} \vartheta_\mu = \vartheta_{-\mu}$, one finds that \eqref{spec:eom.spin} is solved by 
$\vartheta = \vartheta_\mu + i \vartheta_{-\mu}$, with $\mu= ch$. It follows from the standard formula that the conformal dimension of the dual fermionic operators are simply $\Delta_f =\frac12 +h$. The values of $h$ that can be found in the tables in Appendix~\ref{app:tables}. Again, we list the results in Tables~\ref{spec:table.dims.gen} and \ref{spec:table.dims.sp} for the generic case $j\neq l$ and the special case $j=l$, respectively.

\begin{table}[!ht]
\caption{Conformal dimensions and supermultiplet structure in the generic case $j\neq l$. \label{spec:table.dims.gen}}
\[
\begin{array}{|c|c|c|c|}
\hline
\multicolumn{4}{|c|}{\text{$2(\bar{l}+1)$ fermion supermultiplets ($n=-\bar{l}, -\bar{l}+2,\ldots,\bar{l}$)}} \\
\hline
\text{boson/fermion}\#  &  f  &  b*2  &  f \\
\hline
\Delta_n & \Delta_{n0}= \sqrt{\cll+\frac54 +n} &  \Delta_{n0}+\frac12 & \Delta_{n0}+1 \\
\hline
\hline
\multicolumn{4}{|c|}{\text{$2(\bar{l}+1)$ boson supermultiplets}} \\
\hline
\text{boson/fermion}\#  &  b  &  f*2  &  b \\
\hline
\Delta & \Delta_{1}= \sqrt{\cll}+1 &  \Delta_1+\frac12 & \Delta_1+1 \\
\hline
\Delta & \Delta_{2}= \sqrt{\cll}-1 &  \Delta_2+\frac12 & \Delta_2+1 \\
\hline
\end{array}
\]
\end{table}

\begin{table}[!ht]
\caption{Conformal dimensions and supermultiplet structure in the special case $j= l$. \label{spec:table.dims.sp}}
\[
\begin{array}{|c|c|c|c|}
\hline
\multicolumn{4}{|c|}{\text{$4j$ fermion supermultiplets ($n=-2j, -2j+2,\ldots,2j-2$)}} \\
\hline
\text{boson/fermion}\#  &  f  &  b*2  &  f \\
\hline
\Delta_n & \Delta_{n0}= \sqrt{(2j+1)^2+\frac54 +n} & \Delta_{n0}+\frac12 & \Delta_{n0}+1 \\
\hline\hline
\multicolumn{4}{|c|}{\text{2 fermion supermultiplets}} \\
\hline
\text{boson/fermion}\#  &  f  &  b  &  \text{---} \\
\hline
\Delta & 2j+\frac32 & 2j+2 &  \\
\hline
\hline
\multicolumn{4}{|c|}{\text{boson supermultiplets}} \\
\hline
\text{boson/fermion}\#  &  b*(2j+1)  &  f*(4j+2)  &  b*(2j+1) \\
\hline
\Delta & 2j+2 & 2j+\frac52 & 2j+3 \\
\hline
\text{boson/fermion}\#  &  b*(2j+1)  &  f*(4j)  &  b*(2j-1) \\
\hline
\Delta & 2j & 2j+\frac12 & 2j+1 \\
\hline
\end{array}
\]
\end{table}

%% file: sec4.tex
\section{D2-brane fluctuations}
\label{D2flucts}

In this section we consider the bosonic and fermionic fluctuations of the classical $1/3$-BPS D2-brane discussed in Sec. \ref{review}. The procedure that leads to the quadratic action is the same as the one used in Sec.~\ref{D6flucts} for the D6-brane. The notation remains essentially the same, with the following  logical differences due to dimensionality. Generic D2-brane indices are denoted by $a,b=0,1,9$. When the worldvolume is split into $\adstwo \times S^1$, $\alpha,\beta=0,1$ are used for the $\adstwo$ part, while $\mu=9$ refers to the $S^1$ part. Latin indices $i,j=2,3,4,5,6,7,8$ denote the normal directions. 

\subsection{Bosonic fluctuations}

The starting point is, again, the action \eqref{cd:Dp.action.new}. For the D2-brane, there are three terms, the BI term and two CS terms ($C_3$ and $\mathcal{F}\wedge C_1$). Expanding the BI term to quadratic order, one obtains
\begin{align}
\label{D2:BI}
	\sqrt{-\det M_{ab}} &\to \sqrt{-\det g_{ab}} \sin\alpha \Bigg\{ 1 + \cot\alpha\, \chi^\flatind{4}
	-\frac{\cos\alpha}{\sin^2\alpha} \left(\frac12 \epsilon^{\alpha\beta} f_{\alpha\beta}\right)\\
\notag
	&\quad +\frac1{2\sin^2\alpha} \nabla^\alpha \chi^\flatind{i} \nabla_\alpha \chi_\flatind{i}  
	+\frac12 \nabla^\mu \chi^\flatind{i} \nabla_\mu \chi_\flatind{i} \\
\notag 
	&\quad +\frac1{\sin^2\alpha} \left[(\chi^\flatind{2})^2+ (\chi^\flatind{3})^2\right] 
	-\frac12 (\chi^\flatind{4})^2 
	-\frac18 \left[ (\chi^\flatind{5})^2+ (\chi^\flatind{6})^2+ (\chi^\flatind{7})^2+ (\chi^\flatind{8})^2\right] \\
\notag
	&\quad +\frac1{4\sin^4\alpha} f_{\alpha\beta}f^{\alpha\beta} +\frac1{2\sin^2\alpha} f^{\alpha\mu}f_{\alpha\mu}
	- \frac{\cos^2\alpha}{\sin^3\alpha} \chi^\flatind{4} \left(\frac12 \epsilon^{\alpha\beta} f_{\alpha\beta}\right) \Bigg\}~.
\end{align}
Note that the covariant derivative contains the normal bundle connection,
\begin{equation}
\label{D2:cov.der}
	\nabla_a \chi^\flatind{i} = \partial_a \chi^\flatind{i} + A_a{}^\flatind{i}{}_\flatind{j} \chi^\flatind{j}~,
\end{equation} 
which, in contrast to the D6-brane case, has non-zero components
\begin{equation}
\label{D2:A.conn}
	A_{\mu\flatind{56}} = \frac12 \sin^2\frac{\alpha}2~,\qquad 
	A_{\mu\flatind{78}} = -\frac12 \cos^2\frac{\alpha}2~.
\end{equation} 
The only non-zero component of the second fundamental form is 
\begin{equation}
\label{D2:H}
	H_\flatind{4}{}^\mu{}_\mu = -\cot\alpha~.
\end{equation} 
The WZ term with $C_3$ is similar to the $C_7$ term in the D6-brane case, and leads to the following contribution
\begin{equation}
\label{D2:WZ3}
	P[C_3] \to d^3 \xi\sqrt{-\det g_{ab}} \frac32 e^\mu_\flatind{9} 
	\left( \chi^\flatind{2} \nabla_\mu \chi^\flatind{3} - \chi^\flatind{3} \nabla_\mu \chi^\flatind{2} \right)~.
\end{equation}

The $C_1$ WZ term is similar to the $C_5$ term in the D6-brane case, but contains some additional terms,
\begin{align}
\label{D2:WZ1}
	 \cf\wedge P[C_1] \to d^3 \xi\sqrt{-\det g_{ab}} \Bigg\{&\cot\alpha(1-\cos\alpha)   
	+ \cos \alpha\, \chi^\flatind{4} +\frac{1-\cos\alpha}{\sin\alpha} \left(\frac12 \epsilon^{\alpha\beta} f_{\alpha\beta}\right) \\
\notag 
	& + \chi^\flatind{4} \left(\frac12\epsilon^{\alpha\beta} f_{\alpha\beta} \right)
 	+\frac{\cos^2\alpha}{2\sin\alpha} (\chi^\flatind{4})^2 \\
\notag 
	& +\frac12\cos\alpha\, e^{\mu}_\flatind{9} \left( 
 	\chi^\flatind{5} \nabla_\mu \chi^\flatind{6} - \chi^\flatind{6} \nabla_\mu \chi^\flatind{5} 
 	+\chi^\flatind{7} \nabla_\mu \chi^\flatind{8} - \chi^\flatind{8} \nabla_\mu \chi^\flatind{7} \right) \Bigg\}~.
\end{align}

Finally, we sum the three contributions \eqref{D2:BI}, \eqref{D2:WZ3} and \eqref{D2:WZ1}, drop total derivatives
and express the resulting quadratic action in terms of the open string metric, which again rescales the $\adstwo$ part to have radius $\sin\alpha$, 
\begin{equation}
\label{D2:open.string.metric}
	d\tilde{s}^2 = \sin^2\alpha\, g_{\alpha\beta} d\xi^\alpha d\xi^\beta + g_{\mu\nu} d\xi^\mu d\xi^\nu~. 
\end{equation} 
The final  action is:
\begin{align}
\label{D2:action2}
	S_\Dtwo^{B,2} &= -\frac{T_2}{\sin\alpha} \int d^3\xi\, \sqrt{-\det \tilde{g}_{ab}} \Bigg\{ 
	\frac12 \tn^a \chi^\flatind{i} \tn_a \chi^\flatind{i} 
	+\frac1{\sin^2\alpha} \left[ (\chi^\flatind{2})^2 + (\chi^\flatind{3})^2 \right]
	+ \frac3{\sin\alpha} e^\mu_\flatind{9} \chi^\flatind{3} \nabla_\mu \chi^\flatind{2} \\
\notag
	&\quad 
	-\frac18 \left[ (\chi^\flatind{5})^2+ (\chi^\flatind{6})^2+ (\chi^\flatind{7})^2+ (\chi^\flatind{8})^2\right]
	+ \cot\alpha\, e^\mu_\flatind{9} 
		\left(\chi^\flatind{6} \nabla_\mu \chi^\flatind{5} + \chi^\flatind{8} \nabla_\mu \chi^\flatind{7} \right)  \\
\notag
	&\quad
	-\frac{1}{2\sin^2\alpha} (\chi^\flatind{4})^2
	+ \frac14 \tilde{f}^{ab} \tilde{f}_{ab} 
	- \frac{1}{\sin\alpha} \chi^\flatind{4}\left(\frac12\tilde{\epsilon}^{\alpha\beta} f_{\alpha\beta} \right) 
	\Bigg\}~.
\end{align}
Note that, as in the D6 case, there are a number of terms describing a modification of the naive embedding of $\adstwo\subset \adsfour$. The fluctuations $\chi^\flatind{2}$ and $\chi^\flatind{3}$ contain an extra mixing term that arises from the $C_3$ contribution to the WZ action, see Eq.~\ref{D2:WZ3}. In addition, there are mixing terms for the pairs of scalars $(\chi^\flatind{5},\chi^\flatind{6})$ and $(\chi^\flatind{7},\chi^\flatind{8})$, and these pairs of scalars are affected by the non-zero connections in the normal bundle.

\subsection{Fermionic fluctuations}

The construction of the fermionic action for the $\Dtwo$-brane is similar to the $\Dsix$-brane case. We start with Eq.~(17) of \cite{Martucci:2005rb},
\begin{equation}
	S_{\Dtwo}^{(F)} = \frac{T_2}{2} \int d^3 \xi  \e{-\Phi} 
		\sqrt{-\det M_{ab}} \, \bar{\theta} \left(1 - \Gamma_{\Dtwo}\right)
		\left[ (\tilde{M}^{-1})^{ab} \Gamma_{b} D_a - \Delta\right] \theta,
\end{equation}
where $\Gamma_a$ is the pullback of the gamma matrices $\Gamma_m$, the fermionic field $\theta$ is a 10$d$ Majorana spinor, and $\Gamma_{\Dtwo}$ is given by
\begin{equation}
	\Gamma_{\Dtwo} = \frac{1}{\sin \alpha} 
	\left( -\Gamma^{\flatind{019}} \right) 
	\left( 1 + \cos \alpha \Gamma_{(10)} \Gamma^{\flatind{01}} \right)~.
\end{equation}
The pullback of the covariant derivative is again given by  \eqref{D6flucts:D.pullback}. Explicitly, using \eqref{D2:H} and 
\eqref{D2:A.conn}, we have 
\begin{equation}
	\partial_{\alpha} X^m \nabla_m = \nabla_{\alpha}~, \qquad
	\Gamma^{\mu} \partial_{\mu} X^m \nabla_m = \Gamma^{\mu} \nabla_{\mu} 
	+ \frac12 \cot\alpha \Gamma^{\flatind{4}} 
	+ \frac14 \Gamma^{\mu} A_{\flatind{ij}\mu} \Gamma^{\flatind{ij}}~,
\end{equation}
where
\begin{equation}
	A_{\flatind{ij}\mu} \Gamma^{\flatind{ij}}
	= \sin^2\frac{\alpha}2 \Gamma^{\flatind{56}} 
	- \cos^2\frac{\alpha}2 \Gamma^{\flatind{78}}~.
\end{equation}
The $\kappa$-symmetry is fixed by taking $\theta$ to be chiral, which implies that only terms with an odd number of $\Gamma$-matrices survive in the action. 
The result for the fermionic action after a straightforward calculation, expressed in terms of the open string metric \eqref{D2:open.string.metric}, is 
\begin{align}
\label{D2:ferm.action1}
	S_{\Dtwo}^{(F)} &= \frac{T_{2}}{2\sin\alpha} \int d^3 \xi
 	\sqrt{-\det \tilde{g}_{ab}}\, \bar{\theta} 
 	\e{R \Gamma^{\flatind{01}} \Gamma_{(10)}} 
 	\Big\{ \tilde{\Gamma}^{a} \tilde{\nabla}_a \\
\notag
	&\quad	+ \frac1{4\sin\alpha} \left[ 
	\Gamma^{\flatind{569}} - \Gamma^{\flatind{789}} 
	+ \Gamma^{\flatind{239}} \left( 3 - \Gamma^{\flatind{5678}} \right)
	\right] \Big\} \e{R \Gamma^{\flatind{01}} \Gamma_{(10)}} \theta~,
\end{align}
where the spinor rotation parameter $R$ is given by $\sinh 2R = - \cot \alpha$. In what follows, we shall simply work with the rotated spinor,
$\e{R \Gamma^{\flatind{01}} \Gamma_{(10)}} \theta \rightarrow \theta$.

Given the symmetries of our problem, it is convenient to  decompose the 10$d$ Lorentz group as
\be
SO(9,1) \subset SO(2,1) \times SO(2) \times SO(5),
\ee
corresponding to the (0,1,9), (2,3) and (4,5,6,7,8) directions, respectively. A representation  of the 10$d$ gamma matrices compatible with the above  decomposition is
\begin{align}
\notag
	\Gamma^{a} &= \gamma^{a} \otimes \mathbb{I} \otimes \mathbb{I} \otimes 
	\sigma_1~, &&(a = 0,1,9)~, \\
\notag
	\Gamma^{\flatind{i}} &= \mathbb{I} \otimes \tau^{i-1} \otimes 
	\mathbb{I} \otimes \sigma_2~, &&(i = 2,3)~, \\
\label{D2.gamma.decomp}
	\Gamma^{\flatind{j}} &= \mathbb{I} \otimes \tau^{3} \otimes 	
	\lambda^{\flatind{j}} \otimes \sigma_2~, 
	&&(j = 4,5,6,7,8)~, 
\end{align}
where $\sigma^i$ and $\tau^i$ are two sets of Pauli matrices, and $\lambda^i$ are 5$d$ Euclidean $\gamma$-matrices. The representation \eqref{D2.gamma.decomp} is chiral, 
\begin{equation}
	\Gamma_{(10)} = \pm \mathbb{I} \otimes \mathbb{I} \otimes \mathbb{I} 
	\otimes \sigma_3~,
\end{equation}
where the sign depends on the representations of the $SO(2,1)$ and $SO(5)$ Clifford algebras. To be specific, let us choose the $\gamma^a$ such that $\gamma^\flatind{9}=\gamma^\flatind{01}$, \ie
$\gamma^\flatind{019}=1$. 

Hence, under the decomposition \eqref{D2.gamma.decomp}, the 16-component chiral $\theta$ becomes an octet of 2-component 3$d$ spinors. It is useful to decompose this octet into eigenspinors of the three mutually commuting matrices $\tau^3$, $\lambda^\flatind{56}$ and $\lambda^\flatind{78}$,
\begin{equation}
\label{D2:spinor.octet}
	\lambda^\flatind{56} \theta_{abc} = ip \theta_{pqr}~,\quad
	\lambda^\flatind{78} \theta_{abc} = iq \theta_{pqr}~, \quad
	\tau^3 \theta_{pqr} = r \theta_{pqr}~,\quad
	(p,q,r=\pm1)~.
\end{equation}
The action \eqref{D2:ferm.action1} now becomes
\begin{equation}
\label{D2:ferm.action2}
	S_{\Dtwo}^{(F)} = \frac{T_{2}}{2\sin\alpha} \int d^3 \xi
 	\sqrt{-\det \tilde{g}_{ab}}\, \bar{\theta}_{pqr} 
 	\left\{ \tilde{\gamma}^{a} \tilde{\nabla}_a 
	+ \frac{i}{4\sin\alpha} \gamma^\flatind{01} 
	[ p - q + r (3 - pq) ] \right\} \theta_{pqr}~,
\end{equation}
where the sum over the octet is implicit.

\subsection{Spectrum of D2-brane fluctuations}
\label{D2.spectra}

The doublet of scalars $\chi^\flatind{i}$, $i=2,3$, satisfies
\begin{align}
\label{D2:scal.1}
	\left( -\tn_a \tn^a +\frac{2}{\sin^2\alpha} \right) \chi^\flatind{2} 
	-\frac3{\sin\alpha} e^\mu_\flatind{9} \tn_\mu \chi^\flatind{3} &=0~,\\ 
\label{D2:scal.2}
	\left( -\tn_a \tn^a +\frac{2}{\sin^2\alpha} \right) \chi^\flatind{3} 
	+\frac3{\sin\alpha} e^\mu_\flatind{9} \tn_\mu \chi^\flatind{2} &=0~.
\end{align} 
The system is diagonalized by introducing $\chi^\pm = \chi^\flatind{2}\pm i \chi^\flatind{3}$, 
for which \eqref{D2:scal.1} and  \eqref{D2:scal.2} become
\begin{equation}
\label{D2:scal23}
	\left( -\tn_a \tn^a +\frac{2}{\sin^2\alpha} \pm \frac{3i}{\sin\alpha} e^\mu_\flatind{9} \tn_\mu \right) \chi^\pm =0~. 
\end{equation}
Decomposing into the modes on the $S^1$ factor of the D2-brane worldvolume, which are characterized by an integer $n$, \eqref{D2:scal23} gives rise to 
\begin{equation}
\label{D2:scal23.mode}
	\left( \Box - \frac{n^2\mp 3n +2}{\sin^2\alpha} \right) \chi^\pm_n =0~,
\end{equation}
where $\Box = \tilde{g}^{\alpha\beta} \nabla_\alpha \nabla_\beta$. The conformal dimensions of the dual operators are obtained from the standard formula,
\begin{equation}
\label{D2:scal23.dim}
	\Delta^\pm_n = \frac12 +\left| n\mp \frac32 \right|~.
\end{equation}
These are positive integers.

As for the D6-brane the scalar $\chi^\flatind{4}$ couples to the $\adstwo$-components $a_\alpha$ of the vector field. Their field equations are 
\begin{align}
\label{D2:scal.vec.1}
	\left( \tn_a \tn^a +\frac{1}{\sin^2\alpha} \right) \chi^\flatind{4} +\frac1{\sin\alpha} f &= 0~, \\
\label{D2:scal.vec.2}
	 \tn_a (\tn^a a^\alpha -\tn^\alpha a^a) + \frac1{\sin\alpha} \tilde{\epsilon}^{\alpha\beta} \partial_\beta \chi^\flatind{4} &=0~,
\end{align}
where $f$ stands again for $f=\frac12 \tilde{\epsilon}^{\alpha\beta} f_{\alpha\beta}$. Proceeding as in the D6-brane case gives rise to
\begin{equation}
\label{D2:scal.vec.matrix}
	\begin{pmatrix}
	\Box +\tn_\mu \tn^\mu +\frac{1}{\sin^2\alpha} & \frac1{\sin\alpha} \\
	\frac1{\sin\alpha} \Box & \Box +\tn_\mu \tn^\mu 
	\end{pmatrix}
	\begin{pmatrix} \chi^\flatind{4} \\ f \end{pmatrix} =0~.
\end{equation}
Expanding into modes on $S^1$, \eqref{D2:scal.vec.matrix} yields
\begin{equation}
\label{D2:scal.vec.mode}
	\begin{pmatrix}
	\Box +\frac{1-n^2}{\sin^2\alpha} & \frac1{\sin\alpha} \\
	\frac1{\sin\alpha} \Box & \Box - \frac{n^2}{\sin^2\alpha} 
	\end{pmatrix}
	\begin{pmatrix} \chi^\flatind{4}_n \\ f_n \end{pmatrix} =0~.
\end{equation}
To obtain the conformal dimensions of the dual operators, one formally solves the characteristic equation of \eqref{D2:scal.vec.mode} for $\Box$ and translates the two $\adstwo$ mass eigenvalues into the dual conformal dimensions. The result is 
\begin{equation}
\label{D2.scal.vec.dim}
	\Delta_n^\pm = \frac12 +\left| |n| \pm \frac12 \right|~.
\end{equation}

Consider the doublet of scalars $(\chi^\flatind{5},\chi^\flatind{6})$. Their field equations are given by
\begin{align}
\label{D2:scal.5}
	\left( \tn_a \tn^a +\frac14 \right) \chi^\flatind{5}  
	+\cot\alpha\, e^\mu_\flatind{9} \tn_\mu \chi^\flatind{6} &=0~,\\ 
\label{D2:scal.6}
	\left( \tn_a \tn^a +\frac14 \right) \chi^\flatind{6}  
	-\cot\alpha\, e^\mu_\flatind{9} \tn_\mu \chi^\flatind{5} &=0~.
\end{align} 
Remember that the covariant derivative $\nabla_\mu$ contains the normal connection \eqref{D2:A.conn}. Introducing $\chi^\pm= \chi^\flatind{5}\pm i \chi^\flatind{6}$, we diagonalize the covariant derivative
\begin{equation}
\label{D2:scal56.cov.der}
	\nabla_\mu \chi^\pm = \left[ \partial_\mu \pm \frac{i}4 (\cos\alpha-1) \right] \chi^\pm
\end{equation}
and the field equations, which become
\begin{equation}
\label{D2:scal56.eq}
	\left[ \Box +\tilde{g}^{\mu\mu} \partial_\mu^2 \mp \frac{i}{\sin\alpha} e^\mu_\flatind{9} \partial_\mu \right] \chi^\pm =0~.
\end{equation}
After the decomposition into $S^1$ modes and using the standard dimension formula, one obtains the dual operator conformal dimensions
\begin{equation}
\label{D2.scal.56.dim}
	\Delta^\pm_n = \frac12 +\left| n \mp \frac12 \right|~.
\end{equation}
The analysis for the doublet $(\chi^\flatind{7},\chi^\flatind{8})$ proceeds in an identical fashion and yields the same result.

\begin{table}[ht]
\caption{Bosonic Spectrum \label{table:1}}
\[
\begin{array}{|c | c|} 
 \hline
   \text{Doublet} & \Delta_n^{\pm}  \\  
 \hline \hline 
 (\chi^{\underline{2}}_n, \chi^{\underline{3} }_n) & \frac{1}{2} +  | n \mp \frac{3}{2} |  \\  
 (\chi^{\underline{4}}_n, f_n )                    & \frac{1}{2} +  | |n| \pm \frac{1}{2} | \\ 
 (\chi^{\underline{5}}_n, \chi^{\underline{6} }_n) & \frac{1}{2} +  | n \mp \frac{1}{2} |  \\ 
 (\chi^{\underline{7}}_n, \chi^{\underline{8} }_n) & \frac{1}{2} +  | n \mp \frac{1}{2} |  \\ 
 \hline
\end{array}
\]
\end{table}


To obtain the fermionic spectrum, consider the field equations for the octet of 3$d$ spinors arising from the action \eqref{D2:ferm.action2}, in which we split the Dirac operator into the $\adstwo\times S^1$ parts,
\begin{equation}
\label{D2:ferm.eom}
	\left[ \tilde{\gamma}^{\alpha} \tilde{\nabla}_\alpha 
	+ \frac1{\sin\alpha} \gamma^\flatind{01} 
	\left( 2 \partial_\chi +\frac{i}2 D_{pqr} \right) \right] \theta_{pqr}~,
\end{equation}
where 
\begin{equation}
\label{D2:D.def}
	D_{pqr} = \frac12 \left[p-q+r(3-pq)\right]
\end{equation}
takes the odd integer values $D_{pqr}\in (-3,-1,-1,-1,1,1,1,3)$. The $S^1$ dependence is solved by the a simple exponential, 
\begin{equation}
	\theta \sim \e{i\left(n+\frac12\right)\frac{\chi}2}~,
\end{equation}
where $n$ is an integer. (Remember $\chi\in(0,4\pi)$.) Hence, \eqref{D2:ferm.eom} reduces to the form 
\begin{equation}
\label{D2:ferm.eom2}
	\left( \tilde{\gamma}^{\alpha} \tilde{\nabla}_\alpha 
	+ \frac{i\lambda_{npqr}}{\sin\alpha} \gamma^\flatind{01} \right) 		
	\theta_{npqr}~,
\end{equation}
which is familiar from the D6-brane case. The resulting dual conformal dimensions
\begin{equation}
\label{D2:ferm.dims}
  	\Delta_{npqr} = \frac12 + \left|\lambda_{npqr}\right|
\end{equation} 
are positive half-integers ($1/2, 3/2, \cdots$), which nicely complement the bosonic spectrum to fill supersymmetric multiplets. (It may be useful to shift the value of $n$ depending on the value of $D_{pqr}$.)

\begin{table}[ht]
\caption{Fermionic Spectrum\label{table:2}}
\[
\begin{array}{|c| c | c|} 
\hline
  \theta_{npqr} & \lambda_{npqr} & \Delta_n  \\
\hline \hline 
\theta_{n++ +}, \theta_{n-+ +}, \theta_{n - - +} & n + 1 & \frac{1}{2} +  |n + 1|  \\  
\theta_{n -- -}, \theta_{n+ - -}, \theta_{n++-} & n & \frac{1}{2} + | n | \\ 
\theta_{n -+ -} & n - 1& \frac{1}{2} + |n - 1|  \\ 
\theta_{n +- +} & n + 2 & \frac{1}{2} + | n + 2|  \\ 
\hline
\end{array}
\]
\end{table}

%% file: sec5.tex
\section{Comments on supersymmetry and the spectrum}
\label{comments}
The ABJM theory is a three-dimensional Chern-Simons theory with $U(N)\times U(N)$ gauge group. It contains four complex scalar fields $C_I, (I=1,2,3,4)$ in the bifundamental representation $({\bf N}, \bar{\bf N})$, the corresponding complex conjugates in the $(\bar{\bf N}, {\bf N})$ representation, as well as the fermionic superpartners. The gauge fields are governed by a Chern-Simons action
with opposite integer levels for the two gauge groups, $k$ and $-k$ (see \cite{Aharony:2008ug}  for details). The bosonic symmetry subgroups  of this theory are the conformal group in three dimensions $SO(3,2)$ and the R-symmetry group $SU(4)_R\sim SO(6)_R$; these combine into the supergroup $OSp(6|4)$. In the 't~Hooft limit (large $N$ with fixed $N/k$ ratio) the ABJM theory is conjectured to be dual to type IIA string theory on $AdS_4\times \mathbb{CP}^3$. The bosonic subgroups act as isometries of $AdS_4$ and of $\mathbb{CP}^3$.

Let us now discuss the supersymmetric operator whose dual gravity configurations we have studied in this manuscript. To build these type of Wilson loops one considers only one of the gauge fields of the whole $U(N)\times U(N)$ gauge group, we call it  $A_\mu$. We are  mostly guided by the construction of similar operators in ${\cal N}=4$ SYM  but in the absence of adjoint fields one considers the appropriate combination of bi-fundamentals, $C_I$. Namely \cite{Drukker:2008zx,Rey:2008bh,Chen:2008bp}, 

\be
\label{comments:Wloop}
W=\frac{1}{N}{\rm Tr}_R\, {\cal P}\int \left(i A_\mu \dot{x}^\mu +\frac{2\pi}{k} |\dot{x}|\, M^I_J C_I \bar{C}^J\right) ds.
\ee
It was shown in \cite{Drukker:2008zx,Rey:2008bh,Chen:2008bp} that the above operator preserves a $1/6$ of the 24 supercharges when the loop is a straight line or a circle, and the matrix takes the form $M^I_J={\rm diag}\, (1,1,-1,-1)$.  It is worth mentioning that $1/2$ BPS Wilson loops have also been constructed and have a very different pattern of symmetry breaking \cite{Drukker:2009hy}. The Wilson loops \eqref{comments:Wloop} are invariant under an $SL(2,\mathbb{R})\times U(1) \subset SO(3,2)$. The $SL(2,\mathbb{R})$ part of this subgroup is generated by translation along the line $P_0$, dilatation $D$ and a special conformal transformation $K_0$; the $U(1)$ symmetry is generated by rotations around the line, $J_{12}$. Of the R-symmetry, the Wilson loop preserves an $SU(2)\times SU(2) \subset SU(4)$, as follows from the explicit form of the matrix $M^I_J$, which admits $C_1\leftrightarrow C_2$ and $C_3\leftrightarrow C_4$. The classification of $AdS$ superalgebras that are of interest to us was presented in \cite{Gunaydin:1986fe}. One supergroup in that list that contains the bosonic symmetries discussed here is ${\rm OSp}(4|2)$. In the original classification list of \cite{Gunaydin:1986fe}, this is series (i) using the algebra isomorphism $so(4)\sim su(2)\times su(2)$. In appendix~\ref{supergroup} we recall details of the representations of $OSp(4|2)$; in the main text we use a slightly modified notation more akin to our considerations. 

Let us first consider the spectrum of the D2 brane which is given in tables \ref{table:1} and \ref{table:2}. We see that the degeneracies agree precisely with those of the multiplet of $OSp(4|2)$ presented in table \ref{table:sm}.  Here supersymmetry plays a crucial role. Notice that the D2 brane preserves $1/3$ of the 24 bulk supersymmetries. At the level of the multiplet representation we denote the supercharges by  $Q, Q^{\dagger}$; four can be interpreted as creation operators.

\begin{table}[ht]
\caption{Supermultiplet for the D2 brane fluctuations\label{table:sm}}
\[
\begin{array}{|c| c | c| ccccc |} 
 \hline
  \text{Representation}  &  \Delta & (2 p_1 + 1, 2 p_2 + 1)  & \multicolumn{5}{c|}{\text{Degeneracies}} \\
 \hline \hline 
  | \Phi \rangle               & h                     &   (1,1)           & &1 & & & \\ 
  Q^{\dagger} | \Phi \rangle   &  h + \frac{1}{2}      &   (2,2)           & &1 & 3 & & \\ 
  Q^{\dagger} Q^{\dagger}  | \Phi \rangle   &  h + 1   &   (1,3) + (3,1)   & &  & 3 & 3 & \\ 
  Q^{\dagger} Q^{\dagger} Q^{\dagger} | \Phi \rangle   &  h + \frac{3}{2}  &   (2,2) & & & & 3 & 1 \\ 
  Q^{\dagger} Q^{\dagger} Q^{\dagger} Q^{\dagger} | \Phi \rangle  &  h + 2 &   (1,1) & & & &   & 1\\ 
 \hline
\end{array}
\]
\end{table}
There are a total of 16 states in the multiplet: 8 bosons + 8 fermions. The degeneracies follow directly from states being singlets or triplets of the respective $su(2)$ as indicated in the last column of the table. We found it necessary to shift some of the $AdS_2$ quantum numbers to fit in one multiplet, but the spacing was respected. Thus, the spectrum of excitations of the D2 brane falls neatly into long representations of $OSp(4|2)$.

Let us now consider the spectrum of the D6 brane. This configuration is $1/6$ BPS, meaning that there are only four supercharges, two of which can be considered as creation operators in the representation, more precisely, they raise the $AdS_2$ quantum number. Given that these supercharges are a doublet of $Sp(2)$ we obtain generic multiplets of operators with dimensions $(h, h+\frac{1}{2}, h+1)$. This is nicely respected by the values of $h$ that are listed in tables~\ref{spec:table.dims.gen} and \ref{spec:table.dims.sp}, with the exception of two short fermion multiplets. We emphasize that, generically, the dimensions of bosonic operators are not integers. This is a non-trivial result of our calculation. Because all the states in a given row in tables~\ref{spec:table.dims.gen} and \ref{spec:table.dims.sp} have the same values of the $SO(4)$ quantum numbers $(j,l)$, we see that the supercharges are singlets under $SO(4)$ in contrast to the situation for the D2, where the supercharges were vectors under $SO(4)$. In any case, the fact that the spectra for the D6 fluctuations can be organized into supermultiplets is a nice check of our calculation.

%% file: sec6.tex
\section{Conclusions}\label{Sec:Conclusions}

We have computed the spectra of quantum fluctuations of particular embeddings of D6 and D2 branes with electric flux in their worldvolumes  in the background of $AdS_4\times {\mathbb CP}^3$, which is dual to ABJM theory. These brane configurations are expected to be dual to supersymmetric Wilson loops in higher dimensional representations of the gauge group of ABJM theory.

The results represent by themselves interesting progress within a well-defined class of holographic problems. In particular, regardless of the field theory motivation, the general question of semiclassical quantization of certain brane configurations in string theory backgrounds is of great interest. In this respect we have found a peculiar mixing term that are induced by the top, with respect to the worldvolume dimension, RR potential $C_p$ form in the WZ part of the D-brane action.

The construction of supersymmetric field theories in curved spacetimes plays a central role in localization. In this respect, our results provide explicit constructions of supersymmetric field theories living in curved spaces containing an $AdS_2$ factor.  Arguably, the simplest example in this class is provided by the spectrum of excitations of a supersymmetric D3 brane in $AdS_5\times S^5$ which was obtained in \cite{faraggi:2011bb} and later identified as an ${\cal N}=4$ Abelian vector multiplet living in $AdS_2 \times S^2$ in \cite{Buchbinder:2014nia}. The study of supersymmetric field theories on non-compact spaces is an important problem from the field theoretic point of view and presents a, hopefully surmountable, challenge to the program of supersymmetric localization. To first approximation, the supersymmetric field theory describing the quadratic fluctuations constructed here is more similar to the one for D5 brane fluctuations obtained in \cite{Faraggi:2011ge}, which lead to a field theory on $AdS_2 \times S^4$ with non-canonical couplings between the scalars and the Abelian gauge field. In this manuscript, in comparison with \cite{Faraggi:2011ge}, we have found an interesting new mixing term of the embedding that has not been seen before in any of the embeddings in $AdS_5\times S^5$ analyzed in  \cite{faraggi:2011bb,Faraggi:2011ge}. It is worth highlighting that the mixing is intrinsic to brane embeddings; clearly the string, as discussed in \cite{Sakaguchi:2010dg} cannot contain this type of mixing term.

One set of questions that clearly deserves further investigation is the precise classification of all supersymmetric brane configurations with flux on their worldvolume embedded in $AdS_4\times \mathbb{CP}^3$. In particular, there should be other classical solutions corresponding precisely to the $1/2$ BPS configurations where the nature of $\mathbb{CP}^2$ is manifest as a realization of the unbroken $SU(3)$ R-symmetry group.  One particular candidate which we studied preliminarily (but chose not to report on it here) is a D2 brane that wraps $AdS_2\times S^1\subset AdS_4$.  Another configuration is a D6 whose worldvolume contains $\mathbb{CP}^2\subset \mathbb{CP}^3$. We expect to report on  such matters systematically in a future publication.

A logical continuation of our work would be the computation of the one-loop effective actions of the D2 and D6 configurations we considered in this manuscript. In the context of the AdS/CFT correspondence such calculation yields the one-loop correction to the vacuum expectation value of Wilson loops in  the strong 't Hooft coupling limit of ABJM. Indeed, such an effective action computation was undertaken for the fundamental string in \cite{Kim:2012nd} based on the spectrum obtained in \cite{Sakaguchi:2010dg}. Since the results for the fundamental representation, as they currently stand, do not seem to agree with the field theory side, we defer a systematic analysis of the one-loop effective action to a separate publication. It is worth noting that there has been some success in matching the holographic one-loop corrections to field theory results for certain Wilson loops in ABJM \cite{Aguilera-Damia:2014bqa}. On the field theory side, to the best of our knowledge, some of the vacuum expectation values of Wilson loops in higher rank representations have not been systematically studied, although some results for representations with a small number of boxes were reported in \cite{Hatsuda:2013yua}. The configurations we consider here are dual to Wilson loops in representations whose Young tableaux have a number of boxes of the same order as the rank of the gauge group $N$. To the best of our knowledge the expectation values of such Wilson loops have not been systematically computed on the field theory side.  Having the corresponding  exact field theory results will  ultimately provide grounds for a precision holographic comparison between ABJM theory and strings and branes in $AdS_4\times {\mathbb CP}^3$. We hope to report some progress in this direction soon.

%% file: appA.tex
\section{Representations of $\cpspace^n$}
\label{cpn}

Our starting point is a recursion formula for unit $\cpspace^n$ spaces \cite{Cvetic:2000yp}. In that paper, unit $\cpspace^n$ is defined as the $\cpspace^n$ space that arises from the Hopf fibration of a unit $S^{2n+1}$. Hence, unit $\cpspace^1$ is a 2-sphere of radius $\frac12$. 
Let $d\hat{\Sigma}_n$ and $\hat{J}_m = \frac12 d \hat{A}_m$ be the line element and the K\"ahler form of unit $\cpspace^n$, respectively. Then, for any $m$ and $n$, the following formulas hold \cite{Cvetic:2000yp},
\begin{align}
\label{cpn:recursion.unit}
	d\hat{\Sigma}_{m+n+1}^2 &= d\xi^2 + c^2  d\hat{\Sigma}_{m}^2 + s^2 d\hat{\Sigma}_{n}^2 +c^2 s^2 (d\psi + \hat{A}_m -\hat{A}_n)^2~,\\
\label{cpn:recursion.unit.A}
	\hat{A}_{m+n+1} &= c^2 \hat{A}_m + s^2 \hat{A}_n +\frac12 (c^2-s^2) d\psi~,
\end{align}
where $c=\cos\xi$, $s=\sin\xi$, $\xi\in(0,\pi/2)$, $\psi\in(0,2\pi)$. 

In the present paper, we deal with $\cpspace^n$ spaces with line elements $d\Sigma_n= 2d\hat{\Sigma}_n$. Let us call these unit-2 $\cpspace^n$ spaces, because they arise from the Hopf fibration of an $S^{2n+1}$ of radius $2$. Therefore, unit-2 $\cpspace^1$ is 
just a unit $S^2$.
Let $d\Sigma_n= 2d\hat{\Sigma}_n$, $A_n = 2 \hat{A}_n$ and introduce two new angles by $\alpha=2\xi\in(0,\pi)$, $\chi=2\psi\in(0,4\pi)$. In terms of these, \eqref{cpn:recursion.unit} and \eqref{cpn:recursion.unit.A} become
\begin{align}
\label{cpn:recursion}
	d\Sigma_{m+n+1}^2 &= d\alpha^2 + c^2  d\Sigma_{m}^2 + s^2 d\Sigma_{n}^2 +c^2 s^2 (d\chi + A_m -A_n)^2~,\\
\label{cpn:recursion.A}
	A_{m+n+1} &= c^2 A_m + s^2 A_n +\frac12 (c^2-s^2) d\chi~,
\end{align}
where 
\begin{equation}
\label{cpn:cs.def}
	c=\cos\frac{\alpha}2~,\qquad  s=\sin\frac{\alpha}2~.
\end{equation}
The K\"ahler form of unit-2 $\cpspace^n$ is $J_n =4\hat{J}_n= 2 d \hat{A}_n = d A_n$, \ie there is no factor of 2 now. Explicitly, from \eqref{cpn:recursion.A},
\begin{equation}
\label{cpn:recursion.J}
	J_{m+n+1} = c^2 J_m + s^2 J_n - cs d\alpha\wedge (d\chi + A_m - A_n)~.
\end{equation}

With the help of the above formulas we can recursively construct various coordinate systems of unit-2 $\cpspace^n$. One starts with the unit-2 $\cpspace^1$, which is a unit 2-sphere,
\begin{equation}
\label{cpn:cp1} 
	d\Sigma_1^2 = d\Omega^2 = d\vartheta^2 +\sin^2\vartheta d\varphi^2~,\quad 
	A_1=\cos\vartheta d\varphi~,\quad 
	J_1 = -\sin\vartheta d\vartheta \wedge d\varphi~.
\end{equation}
$\cpspace^2$ is obtained for $m=1$, $n=0$,\footnote{The alternative $m=0$, $n=1$ is equivalent by a change of coordinate $\alpha\to \pi-\alpha$.}
\begin{align}
\label{cpn:cp2} 
	d\Sigma_2^2 &= d\alpha^2 + \cos^2\frac{\alpha}2  d\Omega^2 
		+ \cos^2\frac{\alpha}2\sin^2\frac{\alpha}2 (d\chi+ \cos\vartheta d\varphi)^2~,\\
\label{cpn:cp2.A} 	
	A_2 &= \cos^2\frac{\alpha}2 \cos\vartheta d\varphi+\frac12 \cos\alpha d\chi~.
\end{align}
For $\cpspace^3$, one has two choices. One is $m=n=1$, which yields the representation used in \cite{Drukker:2008zx}. 
\begin{align}
\label{cpn:cp3.1} 
	d\Sigma_3^2 &= d\alpha^2 + \cos^2\frac{\alpha}2  d\Omega_1^2 +\sin^2\frac{\alpha}2 d\Omega_2^2 
		+ \cos^2\frac{\alpha}2 \sin^2\frac{\alpha}2 (d\chi + \cos\vartheta_1 d\varphi_1 - \cos\vartheta_2 d\varphi_2)^2~,\\
\label{cpn:cp3.1.A} 	
	A_3 &= \cos^2\frac{\alpha}2 \cos\vartheta_1 d\varphi_1 + \sin^2\frac{\alpha}2 \cos\vartheta_2 d\varphi_2
		+\frac12 \cos\alpha d\chi~.
\end{align}
The other choice is $m=2$, $n=0$, which gives
\begin{align}
\label{cpn:cp3.2} 
	d\Sigma_3^2 &= d\alpha^2 + \cos^2\frac{\alpha}2  d\Sigma_2^2  
		+ \cos^2\frac{\alpha}2 \sin^2\frac{\alpha}2 (d\chi +A_2)^2~,\\
\label{cpn:cp3.2.A} 	
	A_3 &= \cos^2\frac{\alpha}2 A_2 +\frac12 \cos\alpha d\chi~.
\end{align}

As a corollary of the recursion formula with $n=0$ one easily derives the volume of the unit-2 $\cpspace^n$, 
\begin{equation}
\label{cpn:volume}
	V_n = \frac{(4\pi)^n}{n!}~.
\end{equation}

%% file: appB.tex
\section{Representations of $OSp (4|2)$} 
\label{supergroup}
The supergroup $OSp(4|2)$ with bosonic subgroup $Sp(2)$ and $SO(4)$ is the relevant supergroup for the classification of $1/3$ BPS states in ABJM theory, \ie of states that preserve 8 supercharges. 
The representation theory of this supergroup has been discussed in various articles. Some key general remarks on the construction of unitary super ${OSp}(2N|2)$ representations were given, for example,  in \cite{Gunaydin:1986fe}. A dedicated publication to the representations of ${OSp}(4|2)$ appeared, for example, in \cite{Schmitt:1989qz}. The key quantum nubers arise from the following embedding and isomorphism:
\begin{equation}
OSp (4|2, \mathbb{R})  \supset  \; Sp (2, \mathbb{R} ) \times SO(4) 
   \cong \; Sp(2, \mathbb{R}) \times SO(3) \times SO(3)~.
\end{equation}

We can relate the $SO(4)$ labels $(p_1, p_2)$ to  $SO(3) \times SO(3)$ labels $(j, l)$,
\be
j = \frac{1}{2} (p_1 + p_2) , \:\;\;\;\; l = \frac{1}{2} (p_1 - p_2)~.
\ee

The irreducible representations of $OSp (4|2)$ are as follows, with the conditions for the existence of each multiplet given below the corresponding labels (we quote from \cite{Schmitt:1989qz}): 
\begin{align*}
&(h, j, l) \\
& \oplus \underset{2h - j - l \neq 0}{(h + \frac{1}{2} , j + \frac{1}{2} , l + \frac{1}{2})} \oplus 
\underset{l \neq 0}{(h + \frac{1}{2} , j + \frac{1}{2} , l - \frac{1}{2})} \oplus \underset{j \neq 0}{(h + \frac{1}{2} , j - \frac{1}{2} , l + \frac{1}{2})} \oplus \underset{j \neq 0, l \neq 0}{(h + \frac{1}{2} , j - \frac{1}{2} , l - \frac{1}{2})} \\
& \oplus \underset{2h - j - l \neq 0}{(h +1 , j + 1 , l )} \oplus 
\underset{j \neq 0, 2h - j - l \neq 0}{(h +1 , j , l )} \oplus \underset{j \neq 0 , \frac{1}{2}}{(h + 1 , j -1 , l )} \\ 
& \oplus \underset{2h - j - l \neq 0}{(h +1 , j  , l + 1 )} \oplus 
\underset{l \neq 0, 2h + j - l \neq 0}{(h +1 , j , l )} \oplus \underset{l \neq 0 , \frac{1}{2}}{(h + 1 , j  , l - 1 )} \\ 
& \oplus \underset{2h - j - l \neq 0}{(h + \frac{3}{2} , j + \frac{1}{2} , l + \frac{1}{2})} \oplus 
\underset{l \neq 0, 2h + j - l \neq 0}{(h + \frac{3}{2} , j + \frac{1}{2} , l - \frac{1}{2})} \oplus \underset{j \neq 0, 2h - j - l \neq 0}{(h + \frac{3}{2} , j - \frac{1}{2} , l + \frac{1}{2})} \oplus \underset{j \neq 0, l \neq 0}{(h + \frac{3}{2} , j - \frac{1}{2} , l - \frac{1}{2})} \\
& \oplus \underset{ 2h - j - l \neq 0}{(h +2, j,l)} 
\end{align*}

This is the long multiplet in which we accommodated the spectrum of excitations of the D2 brane.

%% file: appC.tex
\section{Harmonic Analysis}
\label{harmonic}

The field equations listed at the end of the previos section involve certain differential operators on the $\Toneone$ part of the $\Dsix$-brane world volume. To deal with these operators, it is appropriate to view $\Toneone$ as a coset manifold \cite{Ceresole:1999zs,Ceresole:1999ht,Benincasa:2011zu}, $\Toneone=\frac{SU(2)\times SU(2)}{U(1)}$, and to apply the powerful technique of \emph{harmonic expansion} \cite{Salam:1981xd}. In this way, their spectrum is obtained in a purely algebraic fashion. The spectrum of Laplace-Beltrami operators on $\Toneone$ was found in \cite{Ceresole:1999zs,Ceresole:1999ht,Benincasa:2011zu}, but the operators arising in our field equations are slightly different. To be self contained, we include a brief review of the geometry of coset manifolds. For a pedagogical introduction to the subject we refer to van~Nieuwenhuizen's lectures \cite{vanNieuwenhuizen:1984ke}. Our signature and curvature conventions agree with those of \cite{vanNieuwenhuizen:1984ke}. In this section, our notation regarding indices is independent of the other sections.

\subsection{Geometry of coset manifolds}

Consider a Lie group $G$ with a subgroup $H$ and their respective Lie algebras $\mathbb{G}$ and $\mathbb{H}$. Decompose $\mathbb{G}$ into 
$\mathbb{G}= \mathbb{H}+ \mathbb{K}$, such that, for the generators $T_a\in \mathbb{K}$ and $T_i\in \mathbb{H}$ and assuming $\mathbb{H}$ to be compact or semi-simple, the structure equations of $\mathbb{G}$ take the form
\begin{equation}
\label{harmonic:algebra}
\begin{aligned}
	\comm{T_i}{T_j} &= C_{ij}{}^k T_k~,\\
	\comm{T_i}{T_a} &= C_{ia}{}^b T_b~,\\
	\comm{T_a}{T_b} &= C_{ab}{}^c T_c + C_{ab}{}^i T_i~.
\end{aligned}
\end{equation}
Starting from any coset representative $L(x)$, define the Lie-algebra valued one-form 
\begin{equation}
\label{harmonic:V.decomp}
	V(x) = L^{-1}(x)dL(x) = r(a) V^a(x) T_a + \Omega^i(x) T_i~.
\end{equation}
Here, $V^a$ are the (rescaled) vielbeins, $r(a)$ denote scale factors, which are independent for each irreducible block of $C_{ia}{}^b$, and
$\Omega^i$ are the $H$-connections. The Maurer-Cartan equation for $V$ yields
\begin{align}
\label{harmonic:MC1}
	dV^a + \frac12 \frac{r(b)r(c)}{r(a)} C_{bc}{}^a V^b\wedge V^c + C_{ib}{}^a \Omega^i \wedge V^b &=0~,\\
\label{harmonic:MC2}
	d\Omega^i + \frac12 r(a)r(b) C_{ab}{}^i V^a\wedge V^b + \frac12 C_{jk}{}^i \Omega^j \wedge \Omega^k &=0~.
\end{align}
Indices will be lowered and raised using a flat coset metric $\eta_{ab}$ and its inverse $\eta^{ab}$, respectively. Later, we shall choose $\eta_{ab}$ to be positive definite Euclidean, but for the time being it is sufficient to state that $\eta^{ab}$ is pseudo-Euclidean with arbitrary signature. 

The geometry of the coset manifold is characterized, as usual, by a torsionless connection defined by 
\begin{equation}
\label{harmonic:connection}
	dV^a + \mathcal{B}^a{}_b \wedge V^b =0~,\qquad \mathcal{B}^{ab}=-\mathcal{B}^{ba}~.
\end{equation}
The Riemann curvature 2-form is
\begin{equation}
\label{harmonic:curvature.form}
	\mathcal{R}^a{}_b = d\mathcal{B}^a{}_b + \mathcal{B}^a{}_c \wedge \mathcal{B}^c{}_b~.
\end{equation}
Comparison of \eqref{harmonic:MC1} and \eqref{harmonic:connection} yields
\begin{equation}
\label{harmonic:Bab}
	\mathcal{B}^a{}_b = \frac12 \mathbb{C}_{cb}{}^a V^c + C_{ib}{}^a \Omega^i~,
\end{equation}
where
\begin{equation}
\label{harmonic:Cabc}
	\mathbb{C}_{cb}{}^a = \frac{r(b)r(c)}{r(a)} C_{cb}{}^a + \frac{r(a)r(c)}{r(b)} C^a{}_{cb} + \frac{r(a)r(b)}{r(c)} C^a{}_{bc}~.
\end{equation}

The $SO(d)$ covariant derivative is defined by 
\begin{equation}
\label{harmonic:cov.diff.def}
	D = d +\frac12 \mathcal{B}^{ab} \mathbb{D}(T_{ab})~,
\end{equation} 
where $\mathbb{D}$ is a representation of $SO(d)$ satisfying
\begin{equation}
\label{harmonic:SOd}
	\comm{\mathbb{D}(T_{ab})}{\mathbb{D}(T_{cd})} = \eta_{bc} \mathbb{D}(T_{ad}) + \eta_{ad} \mathbb{D}(T_{bc})
		- \eta_{ac} \mathbb{D}(T_{bd}) - \eta_{bd} \mathbb{D}(T_{ac})~.
\end{equation} 
A coset harmonic is given, in an arbitrary representation of $G$, by the inverse of a coset representative, 
\begin{equation}
\label{harmonic:Y.def}
	Y(x) = L^{-1}(x)~.
\end{equation}
By definition, it satisfies
\begin{equation}
\label{harmonic:dY}
	dY = -VY=-\left[ r(a) V^a T_a + \Omega^i(x) T_i \right] Y~,
\end{equation}
where the algebra elements act on $Y$ by \emph{right} action. $Y$ also forms a representation of $SO(d)$, if the action of $T_i$ is given by
\begin{equation}
\label{harmonic:Y.SOd}
	\left[ T_i +\frac12 C_i{}^{ab} \mathbb{D}(T_{ab}) \right] Y = 0~.
\end{equation} 
As a consequence, the covariant derivative \eqref{harmonic:cov.diff.def} of an harmonic reduces to
\begin{equation}
\label{harmonic:cov.diff}
	D Y  = V^a D_a Y = - V^a \left[ r(a) T_a + \frac14 \mathbb{C}_a{}^{bc} \mathbb{D}(T_{bc}) \right] Y~. 
\end{equation}  

\subsection{Geometry of $\Toneone$}
\label{harmonic:geometry}

Let us now apply these general results to $\Toneone=\frac{SU(2)\times SU(2)}{U(1)}$. Take $T_1, T_2, T_3$ and $T_{\hat{1}}, T_{\hat{2}}, T_{\hat{3}}$ to be the generators of the first and second $SU(2)$, respectively, let $i=1,2$, $\hat i=\hat 1, \hat 2$, and define 
\begin{equation}
\label{harmonic:TH.def}
	T_5 = T_3 - T_{\hat{3}}~,\qquad T_H = T_3 + T_{\hat{3}}~,
\end{equation}
where $T_H$ generates the $U(1)$. In this basis, the structure equations of $G=SU(2)\times SU(2)$ read
\begin{equation}
\label{harmonic:G.algebra}
\begin{aligned}
	\comm{T_i}{T_j} &= \frac12 \epsilon_{ij} (T_H +T_5)~,\qquad &
	\comm{T_{\hat{i}}}{T_{\hat{j}}} &= \frac12 \epsilon_{\hat{i}\hat{j}} (T_H -T_5)~,\\
	\comm{T_H}{T_i} &= \comm{T_5}{T_i} = \epsilon_i{}^j T_j~, &
	\comm{T_H}{T_{\hat{i}}} &= -\comm{T_5}{T_{\hat{i}}} = \epsilon_{\hat{i}}{}^{\hat{j}} T_{\hat{j}}~.
\end{aligned}
\end{equation}
Defining the scale parameters of the irreducible blocks by 
\begin{equation}
\label{harmonic:abc}
	r(i) = a~,\qquad r(\hat i) =b~,\qquad r(5)=c~,
\end{equation}
the spin connections \eqref{harmonic:Bab} are found as
\begin{equation}
\label{harmonic:Bs}
\begin{aligned}
	B^{5i} &= \frac{a^2}{4c} V^j \epsilon_j{}^i~,\quad &
	B^{ij} &= -\epsilon^{ij} \left[ \omega +\left( c -\frac{a^2}{4c} \right) V^5 \right]~,\\
	B^{5\hat i} &= -\frac{b^2}{4c} V^{\hat j} \epsilon_{\hat j}{}^{\hat i}~,\quad &
	B^{\hat i\hat j} &= -\epsilon^{\hat i \hat j} \left[ \omega -\left( c -\frac{b^2}{4c} \right) V^5 \right]~.
\end{aligned}
\end{equation}
The Ricci tensor $R_{ab} = \mathcal{R}^c{}_{acb}$ turns out to be block-diagonal,
\begin{equation}
\label{harmonic:Ricci}
	R^i{}_j = \delta^i_j \left( a^2 -\frac{a^4}{8c^2} \right)~,\quad
	R^{\hat i}{}_{\hat j} = \delta^{\hat i}_{\hat j} \left( b^2 -\frac{b^4}{8c^2} \right)~,\quad
	R^5{}_5 = \frac{a^4+b^4}{8c^2}~.
\end{equation} 

In is convenient to work in a complex basis, with
\begin{equation}
\label{harmonic:complex.basis}
	x^\pm = \frac12 (x^1 \pm i x^2)~,\qquad x^{\hat\pm} = \frac12 (x^{\hat 1} \pm i x^{\hat 2})~,
\end{equation}
such that the positive definite Euclidean metric $\eta_{ab}$ is given by 
\begin{equation}
\label{harmonic:metric}
	\eta_{+-} = \eta_{\hat + \hat -} = 2~, \qquad \eta_{55} =1~,
\end{equation}
and the components of the $\epsilon$ tensors are 
\begin{equation}
\label{harmonic:eps.tensor}
	\epsilon_{\pm}{}^{\pm} = \epsilon_{\hat{\pm}}{}^{\hat{\pm}} = \pm i~.
\end{equation}
In this basis, the covariant derivatives \eqref{harmonic:cov.diff} are given by
\begin{align}
\label{harmonic:cov.diff.D}
	D_\pm &= -a T_\pm \pm \frac{ia^2}{4c} \mathbb{D}(T_{5\pm})~,\\
\notag
	D_{\hat \pm} &= -b T_{\hat \pm} \mp \frac{ib^2}{4c} \mathbb{D}(T_{5\hat \pm})~,\\
\notag 
	D_5 &= -c T_5 +\frac{i}2 \left( c -\frac{a^2}{4c} \right) \mathbb{D}(T_{+-})
	-\frac{i}2 \left( c -\frac{b^2}{4c} \right) \mathbb{D}(T_{\hat + \hat -})~.
\end{align}
A suitable representation (by right action) of the $SU(2)\times SU(2)$ generators is\footnote{Notice that the role of $T_\pm$ and $T_{\hat \pm}$ as $SU(2)$ raising and lowering operators is the opposite compared to what is indicated by their indices. This is a consequence of right action.}
\begin{align}
\label{harmonic:group.action}
	T_\pm Y^{j,l,r}_q &= -i \left( j \pm \frac{q+r}2\right) Y^{j,l,r\mp1}_{q\mp1}~,\\
\notag
	T_{\hat \pm} Y^{j,l,r}_q &= -i \left( l \pm \frac{q-r}2\right) Y^{j,l,r\pm1}_{q\mp1}~,\\
\notag
	T_5 Y^{j,l,r}_q &= ir Y^{j,l,r}_q~, \\
\notag
	T_H Y^{j,l,r}_q &= iq Y^{j,l,r}_q~.
\end{align}

\subsection{Spectrum of operators on $\Toneone$}
\label{harmonic:spectrum}

We are interested in the spectrum of the differential operators on $\Toneone$, which appear in the field equations listed in subsection~\ref{D6.field.equations}. The scale parameters $a$, $b$ and $c$ are related to the angle $\alpha$ by
\begin{equation}
\label{harmonic:abc.alpha}
	a^2 = \frac1{\cos^2 \frac{\alpha}2}~,\qquad 
	b^2 = \frac1{\sin^2 \frac{\alpha}2}~, \qquad 
	c^2 = \frac1{\sin^2 \alpha}~.
\end{equation}
This leaves a sign ambiguity, which will be resolved shortly. Notice that \eqref{harmonic:abc.alpha} implies
\begin{equation}
\label{harmonic:abc.simp}
	a^2 +b^2 =4c^2~,
\end{equation}
which will simplify many expressions in the sequel.

\paragraph{Scalar fields}	
Scalar fields transform trivially under $SO(d)$, which implies $q=0$ by \eqref{harmonic:Y.SOd}. Vectors (with covariant indices) transform under $\mathbb{D}(T_{ab})_c{}^d=\eta_{ac}\delta_b^d - \eta_{bc}\delta_a^d$. Notice that $D_a Y$ is a vector. We can now calculate the Laplacian $\Box_0=D_a D^a$ of a scalar harmonic, which results in 
\begin{equation}
\label{harmonic:scalar.laplace}
	- \Box_0 Y^{j,l,r}_0 = H_0 Y^{j,l,r}_0~,
\end{equation}
where
\begin{equation}
\label{harmonic:H.def}
	H_0 = a^2 j(j+1) +b^2 l(l+1) -\frac{r^2}4 \left(a^2+b^2-4c^2\right)~.
\end{equation}
This is independent of $r$ by virtue of \eqref{harmonic:abc.simp}. Using \eqref{harmonic:abc.alpha}, let us rewrite it as
\begin{equation}
\label{harmonic:H.rewrite}
	H_0 = c^2 (\cll -1) ~, \qquad \cll = \sin^2\frac{\alpha}2 (2j+1)^2 +\cos^2\frac{\alpha}2 (2l+1)^2~,
\end{equation}
Because of the relations $q=m_3+m_{\hat 3}=0$ and $r=m_3-m_{\hat 3}$, where $m_3$ and $m_{\hat 3}$ are $SU(2)$ quantum numbers, it must hold that $j$ and $l$ are either both integer or half-integer. Accordingly, $r$ is an even or odd integer with $|r|\leq \bar{l}= 2 \min(j,l)$. 

The field equation \eqref{eom:scal} contains, however, the operator
\begin{equation}
\label{harmonic:scal.modified.op}
	-\Box_0' Y  = \left( -\Box_0 \pm i c D_5 \right) Y~,
\end{equation}
where the sign depends on whether $\chi^+$ or $\chi^-$ is considered (and on the still ambiguous sign of $c$). It is straightforward to obtain
\begin{equation}
\label{harmonic:scal.modified.laplace}
	- \Box_0' Y^{j,l,r}_0 = \left(H_0 \pm c^2 r\right) Y^{j,l,r}_0~.
\end{equation}

\paragraph{Vector fields}	
Consider vector fields with covariant indices. The Laplace-Beltrami operator is given by 
\begin{equation}
\label{harmonic:vec.laplace}
	-\Box_1 Y_a = \left( -\delta_a^b D_c D^c +R_a^b \right) Y_b~,
\end{equation}
From \eqref{harmonic:Y.SOd} and \eqref{harmonic:group.action} we deduce that the components of $Y_a$ must carry the follwing quantum numbers,
\begin{equation}
\label{harmonic:Y.vec.nums}
	Y = \begin{pmatrix} Y^{j,l,r\mp1}_{\mp1} \\ Y^{j,l,r\pm1}_{\mp1} \\ Y^{j,l,r}_{0} \end{pmatrix}~.
\end{equation}
After evaluating the covariant derivatives and using \eqref{harmonic:Ricci}, one obtains the matrix form
\begin{equation}
\label{harmonic:vec.laplace.mat}
	-\Box_1 Y_a = 
	\begin{pmatrix} 
		H_0 \pm \frac{a^2}2 r & 0 & \pm \frac{a^3}{4c} (2j\pm r) \\
		0 & H_0 \mp \frac{b^2}2 r & \mp \frac{b^3}{4c} (2l\mp r) \\
		\pm \frac{a^3}{8c} (2j+2 \mp r) & \mp \frac{b^3}{8c} (2l+2 \pm r) & H_0 +\frac{a^4+b^4}{4c^2} 
	\end{pmatrix} Y~.
\end{equation}
We remark that this result corrects some opf the results of Benincasa and Ramallo \cite{Benincasa:2011zu}. In fact, in contrast to what was found in \cite{Benincasa:2011zu}, $H_0$ always is an eigenvalue of this matrix, belonging to the longitudinal vector $D_a Y^{j,l,r}_0$. 

For the field equation \eqref{eom:vec} we need the operator
\begin{equation}
\label{harmonic:vec.modified.op}
	-\Box_1' Y_a  = -\Box_1 Y_a -\cot\alpha\, \calE_a{}^{cb} D_c Y_b~.
\end{equation}
Direct evaluation yields
\begin{equation}
\label{harmonic:vec.Betti.mat}
	\calE_a{}^{cb} D_c Y_b = - 
	\begin{pmatrix} 
		\pm rc & 0 & \pm \frac{a}{2} (2j\pm r) \\
		0 & \pm rc & \pm \frac{b}{2} (2l\mp r) \\
		\pm \frac{a}{4} (2j+2 \mp r) & \pm \frac{b}{4} (2l+2 \pm r) & \frac{a^2-b^2}{2c} 
	\end{pmatrix} Y
\end{equation}
The factor $\cot\alpha$ is determined  \eqref{harmonic:abc.alpha} up to a sign, which is related to the (unfixed) frame orientation. One realizes that the terms in \eqref{harmonic:vec.laplace.mat} and \eqref{harmonic:vec.Betti.mat} combine very nicely (cancelling the asymmetries in $a$ and $b$), if the sign is fixed such that\footnote{In \cite{Ceresole:1999ht}, the sign was fixed imposing supersymmetry on $\Toneone$. In our case $\Toneone$ is not Einstein, so there are no Killing spinors.}
\begin{equation}
\label{harmonic:cot.alpha}
	c = \frac1{\sin\alpha} \qquad \Rightarrow \qquad \cot \alpha = \frac{b^2-a^2}{4c}~.
\end{equation}
Therefore, simplifying also by \eqref{harmonic:abc.simp}, we obtain 
\begin{equation}
\label{harmonic:vec.eom.mat}
	-\Box_1' Y_a = 
	\begin{pmatrix} 
		H_0 \pm r c^2 & 0 & \pm \frac{ac}2 (2j\pm r) \\
		0 & H_0 \mp r c^2 & \mp \frac{bc}2 (2l\mp r) \\
		\pm \frac{ac}{4} (2j+2 \mp r) & \mp \frac{bc}{4} (2l+2 \pm r) & H_0 +2c^2 
	\end{pmatrix} Y~.
\end{equation}
It is straightforward to calculate the eigenvalues of this matrix, but we have to be slightly more detailed 
in the analysis of the spectrum. The fact that each non-zero component of the vector \eqref{harmonic:Y.vec.nums} must be a valid representation of $SU(2)\times SU(2)$ poses a number of restrictions. As for scalar fields, $j$ and $l$ must both be integers or half-integers, with $r$ even or odd, respectively. The restrictions on the range of $r$ that arise from the non-zero vector components are summarized in Tab.~\ref{harmonic:vec.table.r}. The overall range of $r$ for a given eigenvector is obtained as the intersection of all the restrictions, taking care of vanishing vector components. Our results for the eigenvectors, eigenvalues, and ranges of $r$ are listed in Appendix~\ref{app:tables}.

\begin{table}[!ht]
\caption{Restrictions on $r$ for non-zero components of the vector \eqref{harmonic:Y.vec.nums}. \label{harmonic:vec.table.r}}
\[
\begin{array}{|c|c|c|c|}
\hline
\text{component} & SU(2)\times SU(2) \text{ rep.} & \multicolumn{2}{c|}{\text{restrictions on $r$}} \\
\hline
+ & Y^{j,l,r-1}_{-1}  & -2j+2\leq r \leq 2j+2 & -2l\leq r \leq 2l \\
- & Y^{j,l,r+1}_{1} & -2j-2\leq r \leq 2j-2 & -2l\leq r \leq 2l \\
\hat + &  Y^{j,l,r+1}_{-1} & -2j\leq r \leq 2j & -2l-2\leq r \leq 2l-2 \\
\hat - & Y^{j,l,r-1}_{1} & -2j\leq r \leq 2j & -2l+2\leq r \leq 2l+2 \\
5 & Y^{j,l,r}_{0} & -2j \leq r \leq 2j & -2l\leq r \leq 2l \\
\hline
\end{array}
\]
\end{table}

\paragraph{Spinor fields}	
In our conventions, the $SO(d)$ generators acting on spinors are $\mathbb{D}(T_{ab})=\Sigma_{ab} =\frac14[\gamma_a,\gamma_b]$, where the Dirac matrices satisfy $\gamma_a\gamma_b+\gamma_b\gamma_a =2\eta_{ab}$. We choose them as
\begin{equation}
\label{harmonic:Dirac.mats}
	\gamma_i = \sigma_i \times \mathbb{I}~,\qquad 
	\gamma_{\hat i} = \sigma_3 \times \sigma_i~,\qquad
	\gamma_5 = \sigma_3\times \sigma_3~.
\end{equation}
Notice that they satisfy $\gamma_{12\hat 1 \hat 2 5} =-1$. 
Furthermore, in the complex basis \eqref{harmonic:complex.basis}, we have 
\begin{equation}
\label{harmonic:Pauli.mats}
	\sigma_+ = \sigma_1 -i \sigma_2 = \begin{pmatrix} 0&0 \\ 2&0 \end{pmatrix}~,\qquad
	\sigma_- = \sigma_1 +i \sigma_2 = \begin{pmatrix} 0&2 \\ 0&0 \end{pmatrix}~.
\end{equation}
This implies that the $SO(d)$ generators needed in the covariant derivatives \eqref{harmonic:cov.diff.D} are
\begin{equation}
\label{harmonic:spinor.mats}
\begin{aligned}
	\Sigma_{5\pm} &= \mp \frac12 \sigma_\pm \times \sigma_3~,\qquad &
	\Sigma_{5\hat \pm} &= \mp \frac12 \mathbb{I} \times \sigma_\pm~,\\
	\Sigma_{+-} &= - \sigma_3 \times \mathbb{I}~, &
	\Sigma_{\hat +\hat -} &= - \mathbb{I} \times \sigma_3~.
\end{aligned}
\end{equation}
The branching of this representation into representations of $U(1)$ is given by
\begin{equation}
\label{harmonic:spinor.branching}
	-\frac12 C_H{}^{ab} \Sigma_{ab} = i \begin{pmatrix} -1&&&\\ &0&&\\ &&0& \\ &&&1 \end{pmatrix}~.  
\end{equation}

We can now construct the Dirac operator, $\slD=\gamma^a D_a$. Direct evaluation yields
\begin{align}
\notag 
	\slD &= -\frac{a}2 (\sigma_- T_+ + \sigma_+ T_-) \times \mathbb{I} 
			-\frac{b}2 \sigma_3 \times (\sigma_{\hat -} T_{\hat +} + \sigma_{\hat +} T_{\hat -}) 
			- c T_5 \sigma_3 \times \sigma_3 \\
\notag &\quad 
			-\frac{i}2 \left( c +\frac{a^2}{4c} \right) \mathbb{I} \times \sigma_3 
			+\frac{i}2 \left( c +\frac{b^2}{4c} \right) \sigma_3 \times \mathbb{I} \\
\notag
		&= \begin{pmatrix} 
			-cT_5        & -bT_{\hat +} & -a T_+      & 0 \\
			-bT_{\hat -} & cT_5         & 0           & -a T_+ \\
			-a T_-       & 0            & cT_5       & bT_{\hat +} \\
			0            & -aT_-        & bT_{\hat -} & -cT_5 
			\end{pmatrix} \\
\label{harmonic:Dirac.op}
		&\quad + \frac{i}{8c} \begin{pmatrix}
			-(a^2-b^2) &&&\\
			& 8c^2 +a^2+b^2 &&\\
			&& -(8c^2 +a^2 +b^2) & \\
			&&& a^2-b^2
			\end{pmatrix}~.       
\end{align} 
The field equation \eqref{eom:spin} contains the operators \eqref{D6:ferm.D.ops}. They become, in the notation of this section,
\begin{equation}
\label{harmonic:spinor.terms}
	\mathcal{D}_\pm = \slD + \frac{i}4 \cot\alpha\,(\Sigma_{+-}+\Sigma_{\hat + \hat -}) \pm \frac{i}{4\sin\alpha}( \gamma_5+3)~.
\end{equation}
Using \eqref{harmonic:cot.alpha}, the additional terms have the following matrix form, 
\begin{equation}
\label{harmonic:spinor.terms.mat}
	\frac{i}{8c}(a^2-b^2) \begin{pmatrix} 1 &&&\\ & 0 &&\\ && 0 & \\ &&& -1 \end{pmatrix}
	\pm \frac{ic}2 \begin{pmatrix} 2 &&&\\ & 1 &&\\ && 1 & \\ &&& 2 \end{pmatrix}~.
\end{equation}
As for the vector case, we realize that the sign of $c$ implied by \eqref{harmonic:cot.alpha} is such that \eqref{harmonic:spinor.terms.mat} cancels the asymmetries between $a$ and $b$ in the Dirac operator \eqref{harmonic:Dirac.op}. 

By inspection of \eqref{harmonic:Dirac.op}, \eqref{harmonic:spinor.branching} and \eqref{harmonic:group.action}, we can establish that the spinor components must carry the following quantum numbers,
\begin{equation}
\label{harmonic:Y.spin.nums}
	Y = \begin{pmatrix} Y^{j,l,r}_{-1} \\ Y^{j,l,r-1}_{0} \\ Y^{j,l,r+1}_{0} \\ Y^{j,l,r}_{1} \end{pmatrix}~.
\end{equation}
This makes it possible to replace the coset generators in \eqref{harmonic:Dirac.op} by numerical values. Using also \eqref{harmonic:abc.simp} we obtain
\begin{equation}
\label{harmonic:spin.eom.mat}
	\mathcal{D}_\pm Y= 
	\frac{i}2 
	\begin{pmatrix}
		c(-2r\pm 2) & b(2l+1-r)   & a(2j+1+r)   & 0 \\
		b(2l+1+r)   & c(2r+1\pm1) & 0           & a(2j+1+r) \\
		a(2j+1-r)   & 0           & c(2r-1\pm1) & -b(2l+1-r) \\
		0           & a(2j+1-r)   & -b(2l+1+r)  & c(-2r\pm2)   
	\end{pmatrix} Y~.
\end{equation}

We proceed as for the vectors, evaluating first the restrictions the $SU(2)\times SU(2)$ representations of the single spinor components impose. Here, $j$ and $l$ are both integer or half-integer, with $r$ odd or even respectively (vice versa with respect to the scalar and vector cases). The restrictions arising from the non-zero components are listed in Table~\ref{harmonic:spin.table.r}.

\begin{table}[!ht]
\caption{Restrictions on $r$ for non-zero components of the spinor \eqref{harmonic:Y.spin.nums}. \label{harmonic:spin.table.r}}
\[
\begin{array}{|c|c|c|}
\hline
SU(2)\times SU(2) \text{ rep.} & \multicolumn{2}{c|}{\text{restrictions on $r$}} \\
\hline
 Y^{j,l,r}_{-1}  & -2j+1\leq r \leq 2j+1 & -2l-1 \leq r \leq 2l-1 \\
 Y^{j,l,r-1}_{0} & -2j+1\leq r \leq 2j+1 & -2l+1 \leq r \leq 2l+1 \\
 Y^{j,l,r+1}_{0} & -2j-1\leq r \leq 2j-1 & -2l-1\leq r \leq 2l-1 \\
 Y^{j,l,r}_{1} & -2j-1\leq r \leq 2j-1 & -2l+1\leq r \leq 2l+1 \\
\hline
\end{array}
\]
\end{table}

\subsection{Tables of harmonics and eigenvalues}
\label{app:tables}

The following tables list the solutions of the harmonic analysis on $\Toneone$ for the vector and spinor fields. 
One must distinguish the generic case $j\neq l$, from the special case $j=l$, for which $\cll$ simplifies to $\cll=(2j+1)^2$. 
Some of the generic solutions simplify in the special case $j=l$, because common factors can be pulled out of the vectors and spinors.
As a consequence, the associated range of $r$ may be smaller than in the generic case.

As discussed in the main text, $j$ and $l$ are both non-negative integer or half-integer, with $r$ even or odd (odd or even), respectively, for vectors (spinors). We define $\bar{l} = 2\min(j,l)$.

\begin{table}[!ht]
\caption{Eigenvectors and eigenvalues of the modified vector Laplacian, $-\Box_1'$, defined in \eqref{harmonic:vec.modified.op} and given in \eqref{harmonic:vec.eom.mat} in matrix form. Generic case $j\neq l$.}
\[
\begin{array}{|c|c|c|c|}
\hline
-\Box'_1 & \text{eigenvector} & \text{eigenvalue} & \text{range of $r$} \\
\hline
 j\neq l  &
\begin{pmatrix} a(2j+r)\\ a (2j-r) \\ b(2l-r) \\ b(2l+r) \\ -2cr \end{pmatrix} &  
	H_0 & |r|\leq \bar{l} \\ 
\cline{2-4} &
\begin{pmatrix} a(2j+r)(r+h)\\ a (2j-r)(r-h) \\ b(2l-r)(r-h) \\ b(2l+r)(r+h) \\ 2c(h^2-r^2) \end{pmatrix} &  
	\begin{matrix} H_0 + h c^2\\ h = 1\pm \sqrt{\cll} \end{matrix} & |r|\leq \bar{l} \\
\cline{2-4} &
\begin{pmatrix} b(2l+2-r)\\ 0 \\ 0 \\ -a (2j+2-r) \\ 0 \end{pmatrix} &  H_0 +r c^2  & |r-2|\leq \bar{l} \\
\cline{2-4} &
\begin{pmatrix} 0 \\ b(2l+2+r)\\ -a (2j+2+r) \\ 0 \\ 0 \end{pmatrix} &  H_0 -r c^2  & |r+2|\leq \bar{l} \\
\hline
\end{array}
\]
\end{table}

\begin{table}[!ht]
\caption{Eigenvectors and eigenvalues of the modified vector Laplacian, $-\Box_1'$. Special case $j=l$.
Only the $h=2j+2$ solution exists for $j=0$, while the $h=-2j$ solution does not exist for $j=\frac12$.} 
\[
\begin{array}{|c|c|c|c|}
\hline
-\Box'_1 & \text{eigenvector} & \text{eigenvalue} & \text{range of $r$} \\
\hline
 j=l &
\begin{pmatrix} a(2j+r)\\ a (2j-r) \\ b(2j-r) \\ b(2j+r) \\ -2cr \end{pmatrix} &  
	4j(j+1)c^2 & |r|\leq 2j > 0 \\
\cline{2-4} &
\begin{pmatrix} a(2j+r)(r+h)\\ a (2j-r)(r-h) \\ b(2j-r)(r-h) \\ b(2j+r)(r+h) \\ 2c(h^2-r^2) \end{pmatrix} &  
	\begin{matrix} [4j(j+1)+ h]c^2 \\ h= 2j+2 \end{matrix} & |r|\leq 2j \\
\cline{2-4} &
\begin{pmatrix} a \\ -a \\ -b \\ b \\ -2c \end{pmatrix} &  
	\begin{matrix} [4j(j+1)+ h]c^2 \\ h=-2j \end{matrix} & |r|\leq 2j-2 \\
\cline{2-4} &
\begin{pmatrix} b\\ 0 \\ 0 \\ -a \\ 0 \end{pmatrix} &  [ (2j+1)^2 +(r-1)] c^2 & |r-1|\leq 2j-1 \\
\cline{2-4} &
\begin{pmatrix} 0 \\ b \\ -a \\ 0 \\ 0 \end{pmatrix} & [ (2j+1)^2 -(r+1)] c^2  & |r+1|\leq 2j-1 \\
\hline
\end{array}
\]
\end{table}

\begin{table}[!ht]
\caption{Eigenvectors and eigenvalues of the spinor operators $\mathcal{D}_\pm$, defined in \eqref{harmonic:spinor.terms} and given in \eqref{harmonic:spin.eom.mat} in matrix form. The eigenvalue is related to $h$ by $\lambda=ich$. Generic case $j\neq l$.}
\[
\begin{array}{|c|c|c|c|}
\hline
& \text{eigenvector} &  h  & \text{range of $r$} \\
\hline
\mathcal{D}_+ &
\begin{pmatrix} b(2l+1-r)\\ 2c(h+r-1) \\ 0 \\ a(2j+1-r) \end{pmatrix} &  
	1\pm \sqrt{\cll} & |r-1|\leq \bar{l} \\
\cline{2-4}
& \begin{pmatrix} a(2j+r+1) \\ 0 \\ 2c(h+r-1) \\ -b(2l+r+1) \end{pmatrix} &  
	\frac12 \pm \sqrt{\frac14 +\cll -r} & |r+1|\leq \bar{l} \\
\hline
\mathcal{D}_- &
\begin{pmatrix} a(2j+r+1) \\ 0 \\ 2c(h+r+1) \\ -b(2l+r+1) \end{pmatrix} &  
	-1\pm \sqrt{\cll} & |r+1|\leq \bar{l} \\
\cline{2-4}
& \begin{pmatrix} b(2l+1-r)\\ 2c(h+r+1) \\ 0 \\ a(2j+1-r) \end{pmatrix} &  
	-\frac12 \pm \sqrt{\frac14 +\cll +r} & |r-1|\leq \bar{l} \\
\hline
\end{array}
\]
\end{table}

\begin{table}[!ht]
\caption{Eigenvectors and eigenvalues of the spinor operator $\mathcal{D}_+$. The eigenvalue is related to $h$ by $\lambda=ich$. Special case $j=l$. Notice that the range of $r$ depends on the sign in the eigenvalue.}
\[
\begin{array}{|c|c|c|c|}
\hline
& \text{eigenvector} &  h  & \text{range of $r$} \\
\hline
\mathcal{D}_+ &
\begin{pmatrix} b(2j+1-r)\\ 2c(h+r-1) \\ 0 \\ a(2j+1-r) \end{pmatrix} &  
	1\pm (2j+1) & \begin{matrix} |r-1|\leq 2j \,\, (+) \\ |r|\leq 2j-1 \,\, (-) \end{matrix} \\
\cline{2-4} &
\begin{pmatrix} a(2j+r+1) \\ 0 \\ 2c(h+r-1) \\ -b(2j+r+1) \end{pmatrix} &  
\frac12 \pm \sqrt{\frac14 +(2j+1)^2 -r} & \begin{matrix} |r|\leq 2j-1 \,\, (+) \\ |r+1|\leq 2j \,\, (-) \end{matrix} \\
\hline
\mathcal{D}_- &
\begin{pmatrix} a(2j+r+1) \\ 0 \\ 2c(h+r+1) \\ -b(2j+r+1) \end{pmatrix} &  
	-1\pm (2j+1) &  \begin{matrix} |r|\leq 2j-1 \,\, (+) \\ |r+1|\leq 2j \,\, (-) \end{matrix} \\
\cline{2-4}
& \begin{pmatrix} b(2j+1-r)\\ 2c(h+r+1) \\ 0 \\ a(2j+1-r) \end{pmatrix} &  
	-\frac12 \pm \sqrt{\frac14 +(2j+1)^2 +r} & \begin{matrix} |r-1|\leq 2j \,\, (+) \\ |r|\leq 2j-1 \,\, (-) \end{matrix} \\
\hline
\end{array}
\]
\end{table}

%% file: WL_ABJM_arxiv.bbl
\providecommand{\href}[2]{#2}\begingroup\raggedright\endgroup